\documentclass[11pt]{article}
\usepackage{amsmath,amsbsy,multirow}
\usepackage{amsfonts,amssymb,latexsym,bm,cite}
\usepackage[margin=1in]{geometry}
\usepackage{booktabs,rotating}
\usepackage{graphicx,color}
\newcommand{\tR}{\ensuremath{\tilde{R}}}
\newcommand{\pval}{$p$\,-value}
\newcommand{\E}{\ensuremath{\mathbb{E}}}
\newcommand{\R}{\ensuremath{\mathbb{R}}}
\newcommand{\V}{\ensuremath{\mathbb{V}}}

\newcommand{\err}{\ensuremath{q_{2}}}
\newcommand{\mH}{\ensuremath{\mathcal{H}}}
\newcommand{\mN}{\ensuremath{\mathcal{N}(0,1)}}
\newcommand{\sig}{\ensuremath{\sigma}}
\newcommand{\sigmle}{\ensuremath{\sigma_{\hat{\alpha}}}}
\newcommand{\al}{\ensuremath{\alpha}}
\newcommand{\almin}{\ensuremath{\alpha_{\text{min}}}}
\newcommand{\almax}{\ensuremath{\alpha_{\text{max}}}}
\newcommand{\ga}{\ensuremath{\gamma}}
\newcommand{\ka}{\ensuremath{\kappa}}
\newcommand{\ta}{\ensuremath{\theta}}

\newcommand{\hal}{\ensuremath{\hat{\alpha}}}
\newcommand{\pP}{\ensuremath{P_{\text{pre}}}}
\newcommand{\tP}{\ensuremath{P_{\text{sim}}}}
\newcommand{\pS}{\ensuremath{S_{\text{pre}}}}

\newcommand{\hco}{\ensuremath{c_{\text{pre}}}}
\newcommand{\tco}{\ensuremath{c_{\text{sim}}}}
\newcommand{\hka}{\ensuremath{\hat{\kappa}}}

\newcommand{\Tlr}{\ensuremath{T_{\text{LR}}}}
\newcommand{\Twe}{\ensuremath{T_{\text{W}}}}
\newcommand{\Two}{\ensuremath{T_{\text{W2}}}}
\newcommand{\Twt}{\ensuremath{T_{\text{W3}}}}
\newcommand{\Twf}{\ensuremath{T_{\text{W4}}}}
\newcommand{\Tsc}{\ensuremath{T_{\text{S}}}}
\newcommand{\Rlr}{\ensuremath{R_{\text{LR}}}}
\newcommand{\Rwe}{\ensuremath{R_{\text{W}}}}
\newcommand{\Rwo}{\ensuremath{R_{\text{W2}}}}
\newcommand{\Rwt}{\ensuremath{R_{\text{W3}}}}
\newcommand{\Rwf}{\ensuremath{R_{\text{W4}}}}
\newcommand{\Rsc}{\ensuremath{R_{\text{S}}}}

\newcommand{\sgn}{\text{sgn}}

\newcommand{\nc}{\normalcolor}

\begin{document}

\title{Improved Inference for the Signal Significance}
% \title{Improved Inference for the Signal Significance via Edgeworth Approximations}
%\title{Improved Inference for the Asymptotic Distribution of the Signal Fraction}
\author{Igor Volobouev \thanks{Department of Physics \& Astronomy, Texas Tech University} 
\and 
A. Alexandre Trindade, \thanks{Department of Mathematics \& Statistics, Texas Tech University}
}
\maketitle

% \linenumbers

\begin{abstract}
We study the properties of several likelihood-based statistics commonly used in testing
for the presence of a known signal under a mixture model with known
background, but unknown signal fraction. Under the null hypothesis of
no signal, all statistics follow a standard normal distribution in
large samples, but substantial deviations can occur at low sample
sizes. Approximations for respective $p$\,-values are derived to
various orders of accuracy using the methodology of Edgeworth expansions.
Adherence to normality is studied, and the magnitude of deviations is
quantified according to resulting \pval{} inflation or deflation. We
find that approximations to third-order accuracy are generally sufficient
to guarantee $p$\,-values with nominal false positive error rates in the
$5\sig$ range (\pval{}$\ =2.87 \times 10^{-7})$ for the classic Wald, score, and
likelihood ratio (LR) statistics at relatively low samples. Not only
does LR have better adherence to
normality, but it also consistently outperforms all other
statistics in terms of
false negative error rates. The reasons for this are shown to be
connected with high-order cumulant behavior gleaned from fourth order
Edgeworth expansions.
Finally, a conservative procedure is suggested for making
finite sample adjustments while accounting for the look elsewhere effect
with the theory of random fields (a.k.a. the Gross-Vitells method).
%
% Finally, a normalization procedure after
% Edgeworth approximation is suggested as a means to conservatively adjust the global $p$\,-value 
%     in the random fields approach (Gross-Vitells method) to
%    determination of signal significance, while taking into account the look elsewhere effect.
%
\end{abstract}

%%%%%%%%%%%%%%%%%%%%%%%%%%%%%%%%%%%%%%%%%%%%%%%%%%%%%%%%%%%%%%%%%%%%%
\section{Introduction}

In particle physics, a typical search for a new particle or resonance
consists in studying invariant masses of decay products and
identifying a~narrow, localized signal spike on top of a slowly
varying background. If the main purpose of the study is determination
of signal existence, the mass of the particle enters the
analysis as the model nuisance parameter. When it can be expected
that the natural width of the
particle under search is small, one can usually assume that the shape
of the hypothesized signal is Gaussian and that its width is
determined by the detector resolution. On the other hand, if the
expected width is comparable to or substantially larger than the
resolution, it has to be treated as another nuisance parameter
in the search.

Reliable determination of the frequentist statistical
significance of signal evidence, i.e., the \pval{}, 
is crucial for substantiating any claim of discovery.
The \pval{} is derived by estimating
how often the statistical fluctuations in the background alone
can produce an apparent signal excess similar to or larger than that
observed in the data. However, the parameters
of the signal, such as its location and width, are no longer
identifiable in a model which represents pure background.
As the Wilks' theorem~\cite{ref:wilks} does not hold in such situations,
the classical nuisance parameter treatment based on profile likelihood
no longer results in a simple asymptotic
behavior of the \pval{}~\cite{ref:davies-1987}.

In particle physics, the problem of proper accounting for the false
positive signal detection over a wide search area is referred to as
the \emph{look elsewhere effect} (LEE)~\cite{ref:lee-2008}.  While
this effect and its associated ``trial factor'' (i.e., the increase
in the \pval{} in comparison with the case of fixed nuisances) can be
estimated numerically, reliable direct 
determination of the \pval{} by computer simulations is often prohibitively expensive
in terms of required CPU time. Fortunately, an accurate asymptotic
treatment of this problem is possible with the aid of techniques
developed within the theory of random
fields~\cite{ref:adler-1981, ref:adler-taylor-2007}.
The statistic used to test for signal presence is
associated with a~random field in the space formed by nuisance parameters.
The probability that the field maximum exceeds some level depends
on the postulated geometry
of the parameter space and on the field covariance function.
It turns out that, if a sufficiently high level is chosen,
the asymptotic behavior of the \pval{} is determined by a few
coefficients only\footnote{These are the coefficients that enter the asymptotic expansion
  of the Euler characteristic of the field excursion set.},
and full knowledge of the covariance function is not necessary.
These coefficients can be evaluated numerically at a~much reduced 
computational cost in comparison with the brute-force \pval{} determination.
In application to particle physics searches, this approach
has became known as the Gross-Vitells
method~\cite{ref:aslee1-2010, ref:vitells-gross-2011, ref:aslee2-2011}.

Determination of the LEE trial factor by the theory of random
fields relies upon the crucial assumption that the pointwise
distributions of these fields are known. The relevant theory is developed for
centered Gaussian fields with unit variance and a few related fields,
such as~$\chi^2$. However, for finite samples, departures of the
signal testing statistic from normality can be substantial.
To ensure applicability of the LEE theory, 
we must assess how well its assumptions are satisfied and constrain
deviations from these assumptions.
With this purpose in mind, we investigate final sample effects for
a number of statistics that can be used to claim signal discovery.
The specific problem we wish to address is accurate testing for the
presence of a signal fraction, $\al$, in the context of a~model
involving an  additive combination of signal and background
distributions.
% In this context, a~positive value for the estimated
% $\al$ can be used to claim that a~new particle
% has been discovered.
% The frequentist statistical significance
% (i.e., the \pval{})  of such a~discovery is derived from
% the distribution of the estimate under the null hypothesis that no
% signal is present in the data. Proper evaluation of this
% significance taking into account the \emph{look elsewhere
% effect (LEE)}, whereby it is desirable to control for false
% positive signal detection over a wide search area,
% is crucial for substantiating any claim of discovery~\cite{ref:lee-2008}.
In what follows, it will be assumed that both the signal and
background are described by
known probability density functions,  $s(x)$ and $b(x)$ respectively, for some univariate or multivariate
$x$, and that the data are fitted by maximum likelihood
to the (mixture) model density:
\begin{equation}\label{eq:model}
 p(x|\al) = \alpha s(x) + (1-\alpha)b(x).  
\end{equation}
Here, $\alpha$ is the parameter of interest, to be inferred from available data.
For searches
with unknown signal location and/or width, this model
corresponds to a~single point in the nuisance parameter space.
If the observed sample points
$x_1,\ldots, x_n$ consist of independent and identically
distributed realizations from model \eqref{eq:model}, the log-likelihood is
\begin{equation} \label{eq:model-loglihood}
  \ell(\al) = \sum_{i = 1}^n \log p(x_i|\al).
\end{equation}
The maximum likelihood estimator (MLE) $\hal$ is
obtained by maximizing $\ell(\al)$ over all
$\al\in\R$.
It is easily seen that $\ell(\al)$ is strictly concave
under continuous background and signal models, and
thus  $\hal$ is straightforward to locate numerically. 
The overall goal is to produce accurate tests of 
\begin{equation}\label{main-hyp-test}
\mH_0:\ \al=0 \qquad\text{vs.}\qquad \mH_1:\
\al>0. 
\end{equation}
%in the context of log-likelihood expression \eqref{eq:model-loglihood}. 

In the statistics literature, \eqref{eq:model} is known as a \emph{mixture
model}~\cite{ref:mclachlan-peel-2000}. Asymptotically (i.e., for
$n\rightarrow \infty$), $\hat{\alpha}$ is consistent, normal, and
efficient under mild regularity conditions\footnote{It should be pointed out that
in this study we do not impose the $\al\in [0,1]$ constraint. In practice, this
constraint would not lead to a noticeable improvement in the result iterpretability.
At the same time, it would introduce an~unnecessary complication
in various derivations by breaking the likelihood regularity
conditions at the boundaries and by adding a point mass at 0
to the $\hal$ distribution.}.
%
% For reasons of interpretability, it
% is often assumed in these models that $\al\in [0,1]$, thus placing a
% restriction on the parameter space and leading to possible violations of the
% regularity conditions. This leads to a~more complex
% (non-standard) large sample regime for the MLE and test statistics
% constructed from it; e.g., if $\al=0$ as occurs under $\mH_0$, the likelihood ratio statistic
% $\Tlr$ in Table~\ref{tab:defs-of-test-stats}
% would then have an asymptotic distribution
% defined by a 50\% mixture of a $\chi^2_1$ (chi-square with 1 degree
% of freedom) and a point mass at zero~\cite{ref:mclachlan-peel-2000,ref:aslee2-2011}. 
% 
% The assumptions of consistency
% and normality are used in the LEE factor determination by the Gross-Vitells
% method~\cite{ref:aslee1-2010, ref:aslee2-2011, ref:vitells-gross-2011}, whereby the corresponding 
%     LEE factor can be estimated using the theory of random
%     fields~\cite{ref:adler-1981, ref:adler-taylor-2007}.
%
In practice, however, we must maintain confidence in the statements
made for finite $n$. Deviations from normality
can be assessed analytically through so-called ``higher-order''
statistical inference techniques~\cite{ref:severini-2000}.
In this manuscript, we develop higher-order approximations
for model~\eqref{eq:model} with the purpose to both provide such an~assessment and
improve frequentist coverage of the relevant tests. 
As outlined in section~\ref{sec:discuss}, adjustments made
to the local significance of the test statistic can also be
translated into a~subsequent conservative estimate of the global \pval{}.

    If $b(x)$ has nuisance parameters, the likelihood must be profiled over
    these parameters.
Although we do not specifically tackle it
    here, application of the techniques detailed
    in~\cite{ref:severini-2000} is also possible in this situation, leading to suitably
    modified versions of the asymptotic
    expansions we derive in the current paper\footnote{However, the
      required calculations are substantially more challenging.}.  Another
    complication we do not address here is the possibility for the
    sample size $N$ to be random, since in a typical particle physics
    experiment  $N$ is Poisson-distributed. However, the analysis we
    present  here is conditional on a fixed value of $N=n$ when
    forming \eqref{eq:model-loglihood}. 
    
%Consequences for Signal Significance Estimation}
% If $s(x)$ has nuisance parameters, the classical regularity conditions
% ensuring consistency and asymptotic normality of
% the MLE break down, primarily due to parameter non-identifiability
% under $\mH_0$. In this case the (fast) asymptotics-based machinery for
% testing detailed in section~\ref{sec:higher-order} is unavailable, and
% one would  have to resort to (slow) Monte Carlo approximation of the
% null distributions of the relevant statistics. An alternative is to
% employ the Gross-Vitells method mentioned above,
%     which requires that pointwise distributions of the fields must
%     be known. The relevant theory is developed for centered
%     Gaussian fields with unit variance and a few
%     related fields, such as~$\chi^2$. However,
%     finite sample effects lead to violations of these assumptions. As
%     outlined in section~\ref{sec:discuss}, the
%     asymptotic expansions we develop in this paper could be utilized
%     to adjust the local significance of the test statistic, leading
%    to a~subsequent conservative estimate of the global \pval{}.

The remainder of the paper is organized as
follows. Section~\ref{sec:higher-order} introduces the statistics that
will be used to test \eqref{main-hyp-test}, and provides an overview
of the higher-order inference techniques to be used for
approximating the resulting $p\,$-values, that will be presented in
detail in section~\ref{sec:edgeworth}. After describing an example
that poignantly illustrates
the need for \pval{} adjustment in section~\ref{sec:example}, we
  confirm the validity of the proposed approximations via some simulation
  experiments in  section~\ref{sec:simulations}. Finally, section~\ref{sec:power}
undertakes a study of type I and type II errors. We end the paper with
a~discussion.

%%%%%%%%%%%%%%%%%%%%%%%%%%%%%%%%%%%%%%%%%%%%%%%%%%%%%%%%%%%%%%%%%%%%%%%%%%%
\section{Higher-Order Inference Techniques}\label{sec:higher-order}
To set up the notation for what follows,
we denote by $\ell_{i}(\al)=\partial^i\ell/\partial\al^i$
the $i$-th derivative of $\ell(\al)$. Now define $J(\al) = -\ell_{2}(\al)$, and the \textit{expected
  information} number as $I(\al) = \E J(\al)$. Also implicit in the
notation is the argument at which a particular derivative is
calculated, e.g., $J(\hal) =
-\left.\ell_{2}(\al)\right|_{\al=\hal}$, which is the usual definition
of the \textit{observed
  information} number. We assume that the usual
regularity conditions for consistency and asymptotic normality of
$\hal$ are satisfied, e.g., \cite[ch.~3]{ref:severini-2000}. Thus, we
place no restrictions on the parameter space, so that $\al\in\R$.

Parametric statistical inference then seeks a
statistic $T\equiv T(\bm{x})$, a function of the data vector $\bm{x}$, which is
used to formulate a rejection rule for the null
hypothesis $\mH_0$, thereby providing,
in some quantifiable sense, an optimal testing procedure (see, e.g.,
\cite{ref:tsh3-2005} for a detailed exposition relating to the notion
of a \emph{uniformly most powerful}, or UMP, test). 
The three classical test statistics most
commonly used in this context are the
\emph{Likelihood Ratio} (LR), \emph{Wald} (two versions), and \emph{Score}, defined
in Table~\ref{tab:defs-of-test-stats}. It is well known that in lack
of a UMP test, these statistics, and
LR in particular, are generally near-optimal~\cite{ref:shao-2003}. 
%%%%%%%%%%%%%
\begin{table}[tbh]
\caption{Definition of the most common versions of the Likelihood Ratio, Wald, and Score statistics
  for testing the hypothesis in \eqref{main-hyp-test}.}%
\label{tab:defs-of-test-stats}
\centering
\begin{tabular}
[c]{llcc}
\toprule
Method & Statistic    & Value  \\
\midrule
Likelihood Ratio & $\Tlr$  & $2[\ell(\hal)-\ell(0)]$ \\
Wald (Expected) & $\Twe$  & $\hal^2I(0)$ \\
Wald (Observed) & $\Two$  & $\hal^2J(\hal)$ \\
Score           & $\Tsc$  & $\ell_{1}(0)^2/I(0)$ \\
Wald-type 3     & $\Twt$  & $\hal^2/\sig_3^2$ \\
Wald-type 4     & $\Twf$  & $\hal^2/\sig_4^2$ \\
\bottomrule
\end{tabular}
\end{table}
%%%%%%%%%%%%%

Since the asymptotic (large sample) variance of
the MLE $\hal$ is $\sigmle^2=I(\al_0)^{-1}$, where $\al_0$ denotes the
true or hypothesized value of $\al$, the Wald statistics  can use any
of the consistent estimators $I(\al_0)^{-1}$, $I(\hal)^{-1}$, 
$J(\al_0)^{-1}$, or $J(\hal)^{-1}$, when standardizing it (the expected and observed versions listed in
Table~\ref{tab:defs-of-test-stats} being the ones in most
common usage). The two
Wald-type statistics appearing at the bottom of
Table~\ref{tab:defs-of-test-stats}, are variants of $\Two$ that use a shortcut for computing
 $\sigmle^2$ so as to circumvent the need to explicitly 
differentiate $\ell(\al)$. Specifically,
% a second-order Taylor series
% expansion of $-\ell(\al)$ about $\hal$ automatically yields an
% estimate of $J(\hal)$
 an estimate of the $\hal$ uncertainty is obtained by locating the two roots of
 $\ell(\hal)-\ell(\al)=1/2$; this technique enjoying wide
 acceptance in particle physics after its popularization by
 \cite{ref:stat-method-phys-1971, ref:minuit-1975}. Letting
$\almin<\almax$ denote the two solutions in question, define $\sig^{-}=\hal-\almin$ and
$\sig^{+}=\almax-\hal$. Two additional approaches for estimating $\sigmle$ are
then $\sig_3=(\sig^{+}+\sig^{-})/2$, and $\sig_4=\sig^{-}$.

Under $\mH_0$, the LR, Wald, and Score statistics are asymptotically distributed as
$\chi^2_1$, to first order\footnote{This fact remains true for Wald and
Score if any of the following versions of ``information number'' are
used in the definition of the statistic:
$I(0)$, $I(\hal)$, $J(0)$, or $J(\hal)$.}. 
We can write this concisely as
$T\dot\sim\chi^2_1$, valid to $O(n^{-1/2})$. The statement that
$T\dot\sim\chi^2_1$ to $k$-th order, or $O(n^{-k/2})$, informally means that for finite~$n$ a~corrected quantity $(1+O(n^{-k/2}))T+O(n^{-k/2})$ can be
  found which is distributed as $\chi^2_1$, for
    $k=1,2,3,\ldots$. In many situations, it is possible to construct \emph{higher-order}
    approximations ($k\geq 2$)  explicitly. These are arranged in powers of
    $n^{-1/2}$, and give us control
    over the differences between the finite $n$ distribution of a
    statistic and its limiting behavior.

For testing the  one-sided alternative $\mH_1$, the \emph{signed} version
of any of the statistics in  Table~\ref{tab:defs-of-test-stats} (say $T$) can be used, by defining
\begin{equation}\label{sign-stat-r}
R = \sgn(\hal)\sqrt{T}.
\end{equation}
In this case, and under $\mH_0$, $R$ is asymptotically distributed as
a standard normal, $\mN$, to first order, whence the corresponding \pval{}
is $P(\mN>r)$, where
$r=\sgn(\hal)\sqrt{t}$ and $t$ is the observed value of $T$ calculated from the sample on
hand. In the ensuing discussion we reserve
the symbols $\Phi(\cdot)$ and $\phi(\cdot)$ for the cumulative distribution
and probability density functions, respectively, of a $\mN$ distribution.

Tools for developing higher-order asymptotic theory
include Taylor series expansions of $\ell(\cdot)$ near $\alpha = 0$, joint
cumulants for the derivatives of $\ell(\cdot)$, and
 Edgeworth approximations to distributions. To any finite order, relationships between cumulants and moments can be determined routinely
 via a~symbolic algebra system (starting from the~Taylor expansion of
 the cumulant generating function). To obtain Edgeworth approximations to
 the probabilities of a given statistic,
    its cumulants are expressed in terms of
    the joint cumulants of the log-likelihood derivatives. The essence
    of these approximations (for a random variable $Z$ whose distribution is
    close to standard normal) is to construct an~approximate density with the  Gram-Charlier expansion
\begin{equation}\label{eq:Gram–Charlier-Exp}
  f(z) = \phi(z) \left(1 + \sum_{j=1}^{\infty} \beta_j H_j(z) \right),
\end{equation}
whereby the coefficients $\beta_j$ are chosen to match the cumulants $\kappa_j$
of the approximated distribution.  The corresponding cumulative
distribution function (CDF) for $f(z)$, $F(z)=\int_{-\infty}^{z}f(x)dx$, is easily found using the following
property of the Hermite polynomials\footnote{For example: $H_2(z)=z^2-1$, $H_3(z)=z^3-3z$, and $H_5(z)=z^5-10z^3+15z$.} in \eqref{eq:Gram–Charlier-Exp}, $\int_{-\infty}^{z} H_j(x)
e^{-x^2/2} dx = -H_{j-1}(z) e^{-z^2/2}$, whence
\begin{equation}\label{eq:CDF-Gram–Charlier-Exp}
F(z) = \Phi(z) - \phi(z)\sum_{j=1}^{\infty} \beta_j H_{j-1}(z) = 1-S(z),
\end{equation}
where $S(z)$ is the so-called \emph{survival function} (or right-tail probability).
The final Edgeworth expansion is now obtained from \eqref{eq:CDF-Gram–Charlier-Exp} by collecting terms in  powers of~$n^{-1/2}$:
\begin{multline}\label{eq:statexpansion}
      F(z) =  \Phi(z) - \phi(z) \bigg[ \kappa_1 + \frac{1}{6} \kappa_3 H_2(z) + \frac{1}{2} (\kappa_1^2 + \kappa_2 - 1) z \\
        + \left( \frac{1}{6} \kappa_1 \kappa_3 + \frac{1}{24} \kappa_4 \right) H_3(z) + \frac{1}{72} \kappa_3^2 H_5(z) + O(n^{-3/2}) \bigg].
\end{multline}
      This expansion assumes the following (typical) cumulant behavior for $Z$:
    $\kappa_1=O(n^{-1/2})$, $\kappa_2 = 1 + O(n^{-1})$, and
    $\kappa_m=O(n^{-(m - 2)/2})$ for all~$m > 2$. It then follows that
    the first four cumulants of $F(z)$ are $\kappa_1$, $\kappa_2$,
    $\kappa_3 + O(n^{-3/2})$, and $\kappa_4 + O(n^{-3/2})$, with
    all subsequent cumulants being $O(n^{-3/2})$ or smaller. With the
    exception of $\Rwf$, all 
    signed versions of the statistics in Table~\ref{tab:defs-of-test-stats} exhibit
    this cumulant behavior. The appropriate version of
    \eqref{eq:statexpansion} for the case in which $\kappa_2 = 1 + O(n^{-1/2})$
    is given in Appendix~\ref{app:torder-w3w4}.

The fundamental
 strategies in seeking improved inference in the
 finite $n$ situation fall into
 the following categories.
\begin{description}
\item[Strategy 1.] More accurate approximations to tail probabilities of $R$ in
  the form:
\[
P(R\leq r) = \Phi(r) + \text{correction} + \text{error}, 
\]
where the ``correction'' is a function of both $n$ and $r$, and the
``error'' is $O(n^{-1})$ or smaller. These are accomplished primarily
through Edgeworth expansions. Throughout the paper we refer to
the resulting tail probabilities as Edgeworth ``approximations'' or ``predictions''.

\item[Strategy 2.] Modifications to $R$, so that $R\mapsto \tilde{R}$, with $\tilde{R}$
  more closely following a $\mN$:
\[
P(\tilde{R}\leq r) = \Phi(r) + \text{error}, 
\]
and the ``error'' is $O(n^{-1})$ or smaller. The challenge in these
modifications is to correct towards normality, while
still retaining the essence (and near-optimal properties) of $R$.
\end{description}
These strategies apply also to $T=R^2$; the only change being that
$\Phi(\cdot)\mapsto 2\Phi(\cdot)-1$, corresponding to a $\chi^2_1$
tail probability.  Approaches under both strategies were gradually developed over the past century, with some
of the
most important contributions emerging from the pioneering work of
\cite{ref:bn-cox-1994}. A~more recent and updated treatment of the
methodology is given by \cite{ref:severini-2000}. In the next section
we detail the explicit calculations involved.

%%%%%%%%%%%%%%%%%%%%%%%%%%%%%%%%%%%%%%%%%%%%%%%%%%%%%%%%%%%%%%%%%%%%%%%%%%%
\section{Approximated $\bm{p}$\,-Values and Normalizing Transformations}\label{sec:edgeworth}
To concisely compute the approximate $p$\,-values for the statistics
% in Table~\ref{tab:defs-of-test-stats}
under study, we define two versions of the
expectation operator: $\E$ and $\E_s$ will denote expectation under $\mH_0$
and under the signal, respectively:
\[ \E[q] := \int q(x) b(x) dx, \qquad
 \E_s[q] := \int q(x) s(x) dx. \]
Additionally, with the notation $V_i:=\E\left[ \ell_i(0)\right]$,
we find that the following (dimensionless and location-scale invariant)
quantities play a key role in the expressions below:
\begin{eqnarray}
    \gamma &=& \frac{V_3}{2(-V_2)^{3/2}} = \frac{\E_s\left[ \frac{s^2}{b^2} \right] - 3 \E_s\left[ \frac{s}{b} \right] + 2 }{\left( \E_s\left[ \frac{s}{b} \right] - 1 \right)^{3/2}}, \label{eq:our-gamma}\\
    \rho &=& -\frac{V_4}{6V_2^2} = \frac{\E_s\left[ \frac{s^3}{b^3} \right] - 4 \E_s\left[ \frac{s^2}{b^2} \right] + 6 \E_s\left[ \frac{s}{b} \right] -  3}{\left( \E_s\left[ \frac{s}{b} \right] - 1 \right)^2}.\label{eq:our-rho}
\end{eqnarray}

We now give, under Strategy 1,  explicit higher-order expansions for the signed versions of the  statistics from Table~\ref{tab:defs-of-test-stats}. The bulk of the work involves computing 
joint cumulants for the derivatives of $\ell(\cdot)$; these forming
the basis for approximating the cumulants that are then substituted  into
\eqref{eq:statexpansion}. Specifically, and following \cite[ch.~5]{ref:severini-2000}, in what follows we denote by
$n\nu_{ijkl}$ the $(i,j,k,l)$-th cumulant\footnote{An abbreviated
  notation will be adhered to by omitting trailing zeros, e.g.,
  $\nu_{1020}\equiv\nu_{102}$.} of
the first four derivatives of $\ell(\al)$ evaluated at $\al=0$,
$\{\ell_{1}(0),\ldots,\ell_{4}(0)\}$. The calculation of these
cumulants is detailed in Appendix~\ref{app:cumulant-calcs}. Proposed
approaches to improving inference under the guise of Strategy 2 are
also briefly discussed.

%%%%%%%%%%%%%%%%%%%%%
\subsection{Strategy 1: approximations to $\bm{p}$\,-values}\label{subsec:strategy-one}
For testing $\mH_0$ vs.~$\mH_1$ under the paradigm of Strategy 1, we compute higher-order
Edgeworth expansions for the tail probabilities of signed versions of the statistics in
Table~\ref{tab:defs-of-test-stats}. 
First, starting from the approximated $\Rlr$ cumulants
given in \cite[sec.~5.4]{ref:severini-2000}, straightforward
calculations give the corresponding approximated cumulants for model
\eqref{eq:model-loglihood} listed in
the first row of Table \ref{tab:cums-of-test-stats}. Substituting these expressions into
\eqref{eq:statexpansion} gives immediately, 
\begin{equation}\label{eq:Rlr-tail-prob}
P(\Rlr\leq z) = \Phi(z)-\phi(z)\left[
  \left(-\frac{\gamma}{6}\right)n^{-1/2} +
  \left(\frac{(3\rho-2\gamma^2)z}{12}\right)n^{-1} +
  O(n^{-3/2})\right].
\end{equation}

Next, we give analogous results for the expected and observed versions
of Wald. This entails first approximating the appropriate cumulants,
expressions for which are given in \cite[sec.~5.3]{ref:severini-2000} to third order
accuracy for $\Rwe$. The first four cumulants appear to be correctly
stated except for the 2nd, the correct version of which should be:
\[
n[\hka_2(\Rwe)-1] =\left( 2\nu_{21}+ 3\nu_{101}+ 3 \nu_{02} + \nu_{0001}\right)\nu_{2}^{-2}
 + \left(\nu_{001}\nu_{3} +\frac{7\nu_{001}^2}{2}\nc + 5\nu_{11}^2
 + 11\nu_{11}\nu_{001} \right)\nu_{2}^{-3}.
\]
Straightforward calculations then give the values in
the second row of Table \ref{tab:cums-of-test-stats}. Substitution of
these into
\eqref{eq:statexpansion} gives eventually
\begin{multline}\label{eq:Rwe-tail-prob}
P(\Rwe\leq z) = \Phi(z)-\phi(z)\left[
  \left(\frac{\gamma H_2(z)}{6}\right)n^{-1/2} \right. + \\
  \left.\left(\frac{(\rho-\gamma^2-1)z}{2}+\frac{(\rho-3)H_3(z)}{24}+\frac{\gamma^2H_5(z)}{72}\right)n^{-1} +
  O(n^{-3/2})\right].
\end{multline}

For the observed version of Wald,
\cite[sec.~5.3]{ref:severini-2000} carries only second order
accuracy in the approximations to the cumulants of $\Rwo$. Working
from first principles (using Taylor series expansions of
$\ell(\cdot)$ with an appropriate number of terms) and following the
general procedure outlined in \cite[ch.~5]{ref:severini-2000} for
Edgeworth expansions, we
obtain the following third-order accurate expressions for the
first four cumulants of $\Rwo$:
\[ 
\hka_1(\Rwo) = \frac{\nu_{11}}{2\nu_2^{3/2}\sqrt{n}}, \qquad \hka_3(\Rwo) = -\frac{\nu_{001}}{\nu_2^{3/2}\sqrt{n}},
\]
\[
n[\hka_2(\Rwo)-1] = \left( 
\nu_{21} + \nu_{02} - \frac{\nu_{0001}}{2}
\right)\nu_2^{-2} 
+ \left(
\frac{7\nu_{11}^2}{4} - \frac{3\nu_{001}^2}{4}
\right)\nu_2^{-3},
\]
and
\[
n\hka_4(\Rwo) = \left( 
\nu_{4} + 6\nu_{21} + 3\nu_{02} - 2\nu_{0001}
\right)\nu_2^{-2} 
+ \left(
6\nu_{11}\nu_{3} + 18\nu_{11}^2 - 3\nu_{001}^2
\right)\nu_2^{-3}.
\]
The conversion of these expressions
in terms of our log-likelihood function \eqref{eq:model-loglihood} is listed in
 Table \ref{tab:cums-of-test-stats}. Substitution of
these values into
\eqref{eq:statexpansion} gives 
\begin{multline}\label{eq:Rwo-tail-prob}
P(\Rwo\leq z) = \Phi(z)-\phi(z)\left[
  -\left(\frac{3\gamma+2\gamma H_2(z)}{6}\right)n^{-1/2} \right. + \\
  \left.\left(\frac{(3\rho-\gamma^2)z}{2}+\frac{(5\rho+2\gamma^2)H_3(z)}{12}+\frac{\gamma^2H_5(z)}{18}\right)n^{-1} +
  O(n^{-3/2})\right].
\end{multline}

Following the same procedure as for $\Rwo$, we
obtain analogous third-order accurate expressions for the
first four cumulants of $\Rsc$:
\[ 
\hka_1(\Rsc) = 0, \qquad
\hka_2(\Rsc) = 1, \qquad
\hka_3(\Rsc) = \frac{\nu_3}{\nu_2^{3/2}\sqrt{n}}, \qquad
\hka_4(\Rsc) = \frac{\nu_4}{n\nu_2^{2}}.
\]
The conversion of these expressions
into our log-likelihood function appears in Table \ref{tab:cums-of-test-stats}.
Substitution of these $\Rsc$ cumulants into
\eqref{eq:statexpansion} gives 
\begin{multline}\label{eq:Rsc-tail-prob}
P(\Rsc\leq z) = \Phi(z)-\phi(z)\left[
  \left(\frac{\gamma H_2(z)}{6}\right)n^{-1/2} + 
\left( \frac{(\rho-3)H_3(z)}{24}+\frac{\gamma^2H_5(z)}{72}\right)n^{-1} +
  O(n^{-3/2})\right].
\end{multline}
Finally, Table \ref{tab:cums-of-test-stats}
also includes $O(n^{-3/2})$ cumulants for
$\Rwt$ and $\Rwf$, obtained by similar calculations.
The tail probabilities for these
statistics are given in Appendix~\ref{app:torder-w3w4}.

% , along with $\Rwt$ and
% $\Rwf$, also in Table \ref{tab:cums-of-test-stats}.
% Similar expressions for the Edgeworth-approximated $\Rwt$ and
% $\Rwf$ tail probabilities are given in Appendix~\ref{app:torder-w3w4}.
%%%%%%%%%%%%%
\begin{table}[tbh]
\caption{Approximations to the first four cumulants of the signed
  versions of the  statistics in Table~\ref{tab:defs-of-test-stats}, in the context of log-likelihood expression
  \eqref{eq:model-loglihood}. The error in these
  approximations is $O(n^{-3/2})$.}%
\label{tab:cums-of-test-stats}
\centering
\begin{tabular}
[c]{lcccc}
\toprule
Statistic & $\hka_1$  & $\hka_2$  & $\hka_3$  & $\hka_4$  \\
\midrule
$\Rlr$    & $-\frac{\gamma}{6\sqrt{n}}$ & $1+\frac{18\rho-13\gamma^2}{36n}$ & $0$ & $0$ \\
$\Rwe$    & $0$ & $1+\frac{\rho-\gamma^2-1}{n}$ &
$\frac{\gamma}{\sqrt{n}}$ & $\frac{\rho-3}{n}$\\
$\Rwo$    & $-\frac{\gamma}{2\sqrt{n}}$ &
$1+\frac{12\rho-5\gamma^2}{4n}$ & $-\frac{2\gamma}{\sqrt{n}}$ &
$\frac{10\rho}{n}$ \\
$\Rsc$    & $0$ & $1$ & $\frac{\gamma}{\sqrt{n}}$ & $\frac{\rho - 3}{n}$ \\
$\Rwt$    & $-\frac{\gamma}{2\sqrt{n}}$ & $1 + \frac{126 \rho -
  65 \gamma^2}{36 n}$ & $-\frac{2 \gamma}{\sqrt{n}}$ & $\frac{10\rho}{n}$ \\
$\Rwf$    & $-\frac{\gamma}{2\sqrt{n}}- \frac{3\rho - \gamma^2}{3n}$ & $1 + \frac{2 \gamma}{3
  \sqrt{n}} + \frac{126 \rho - 53 \gamma^2}{36 n}$ & $-\frac{2
  \gamma}{\sqrt{n}} - \frac{4 \rho - \gamma^2}{n}$ & $\frac{10
  \rho}{n}$ \\
\bottomrule
\end{tabular}
\end{table}
%%%%%%%%%%%%%

%%%%%%%%%%%%%%%%%%%%%%%%
%\newpage
\subsection{Strategy 2: normalizing transformations}
%%%%%%%%%%%%%%%%%%%%%%%
A general transformation for ``normalizing'' the distribution of
    a~statistic $R$ that is already approximately normal, is
\begin{equation}\label{eq:norm-transform}
    \tR(r) = \Phi^{-1}(P(R \leq r)),
\end{equation}
where $r$ denotes the observed value of $R$ computed from the sample
at hand. The key idea is that by invoking an accurate (higher-order)
approximation to the cumulative probabilities that comprise
the argument of the standard normal quantile function $\Phi^{-1}(\cdot)$, the
\emph{probability integral transform} will then ensure better
compliance with a Gaussian distribution. For typical
expansions considered in this study,
\begin{equation}\label{eq:general-edge-cdf}
P(R \leq r)=\Phi(r)-\phi(r)\left[ \frac{a(r)}{\sqrt{n}}+\frac{b(r)}{n} +
  O(n^{-3/2})\right], 
\end{equation}
numerically stable evaluation of~\eqref{eq:norm-transform}
to $O(n^{-3/2})$ can be performed at large $r$ by calculating
\begin{equation}\label{eq:stable-norm-transform}
\tR(r) = S_{\Phi}^{-1}\left(S_{\Phi}(r) + \phi(r)\left[ \frac{a(r)}{\sqrt{n}}+\frac{b(r)}{n}
  \right]\right),
\end{equation}
where $S_{\Phi}(z) \equiv 1-\Phi(z)$ is the survival function of a $\mN$.
This technique, in essence a
combination of Strategies 1 and 2,  would therefore be immediately
applicable to any of the statistics
from Table~\ref{tab:cums-of-test-stats}.

On the other hand, \cite[ch.~7]{ref:severini-2000} discusses at length
a classical
modification to the signed LR statistic
\[ \Rlr(\al)=\sgn(\hal)\sqrt{2[\ell(\hal)-\ell(\al)]}=\sgn(\hal)\sqrt{2\sum_{i=1}^n\log\left(\frac{p(x_i|\hal)}{p(x_i|\al)}\right)}, \]
so that it follows a standard normal
distribution to third-order accuracy. This gives rise to the so-called
Barndorff-Nielsen $R^*$ formula~\cite[sec.~6.6]{ref:bn-cox-1994}:
\begin{equation}\label{eq:r-star-formula}
\Rlr^*(\al) = \Rlr(\al)+\frac{1}{\Rlr(\al)}\log\left|\frac{U(\al)}{\Rlr(\al)}\right|,
\end{equation} 
where $U(\al)$ is a quantity constructed from \emph{sample-space
  derivatives} of $\ell(\al)$. The key idea behind this result is
that for a random variable $X$ with a density of the form
$h(x/\sqrt{n})\phi(x)$, where $h\sim O(1)$, a normalizing
transformation is: $X^*=X-X^{-1}\log h(X/\sqrt{n})$. 

Explicit evaluation of $U(\al)$ is not possible for cases (such as
ours) in which $\ell(\al)$ cannot be expressed in terms of the MLE, but an approximation  based on covariances can be
constructed~\cite[sec.~7.5]{ref:severini-2000}.
However, this involves the
computation of analytically intractable integrals. Another shortcoming
of the $R^{*}$ formula is that it is only available for the LR
statistic. Thus, if $R$ represents any of the statistics in Table
\ref{tab:cums-of-test-stats}, each of which possesses an Edgeworth
expansion of the form \eqref{eq:general-edge-cdf} as shown under Strategy
1, we advocate usage of the normalized statistic \eqref{eq:stable-norm-transform}.
% \begin{equation}\label{eq:our-rprime}
% \tR = \Phi^{-1}\left( \Phi(r)-\phi(r)\left[\frac{a(r)}{\sqrt{n}}+\frac{b(r)}{n}\right]\right).
% \end{equation} 
The resulting $\tR$ will follow a $\mN$ to an accuracy of
$O(n^{-3/2})$ under $\mH_0$. However, it must be kept in mind that
this is an asymptotic statement regarding the order of the error as
$n$ grows. For a given $n$ the magnitude of the error will still be
governed by an appropriate constant that could in some cases be quite large.

To quantify the deviations of $R$ from normality as a function of $r$,
we investigate the quantity
\begin{equation}\label{eq:Delta-R}
    \Delta R(r) = r - \tR(r).
\end{equation}
In combination with \eqref{eq:general-edge-cdf},
the first order Taylor expansion of~\eqref{eq:norm-transform}
yields a simple approximation $\Delta R(r) \approx \frac{a(r)}{\sqrt{n}}+\frac{b(r)}{n}$,
valid to $O(n^{-1})$.

Without loss of generality, for the rest of the paper we will focus on the properties of
the un-normalized $R$, but in an actual application it would behoove the practitioner 
to use the normalized $\tR$ version since computation of
required quantiles for hypothesis testing will be
straightforward. Important results such as  type I and type II error rates
studied in Section \ref{sec:power} for $R$
will be identical for~$\tR$.

%%%%%%%%%%%%%%%%%%%%%%%%%%%%%%%%%%%%%%%%%%%%%%%%%%%%%%%%%%%%%%%%%%%%%%%%%%%
\section{An Illustrative Example}\label{sec:example}
In this section we choose a particular model in order to illustrate the
consequences of applying the Edgeworth expansions for the statistics
in Table \ref{tab:cums-of-test-stats}.
% after appropriate normalization via \eqref{eq:our-rprime}. 
%Thus $\tilde{R}_{\text{LR}}$ will denote the normalized $\Rlr$, etc. 
As choices for the
background and signal under model \eqref{eq:model}, we select a
relatively simple configuration by letting $b(x)$ follow a uniform
distribution on $[0,1]$, and $s(x)$ a truncated Gaussian on $[0,1]$:
\begin{equation}\label{eq:back-signal-model}
  b(x) = \left\{ \begin{array}{ll}
      1, & \mbox{ if } x \in [0, 1] \\
      0, & \mbox{ if } x \not\in [0, 1]
      \end{array} \right.,
\qquad
  s(x) = \left\{ \begin{array}{cl}
      \left. e^{-\frac{(x - \mu)^2}{2 \sigma^2}} \middle/ \int_0^1 e^{-\frac{(y - \mu)^2}{2 \sigma^2}} dy \right., & \mbox{ if }  x \in [0, 1] \\
      0, & \mbox{ if } x \not\in [0, 1]
      \end{array} \right..
\end{equation}
This gives a flat background superimposed with a Gaussian signal. For
this and the remaining sections, whenever specific settings of the
signal are needed, we use 
\begin{equation}\label{eq:back-signal-sets}
\mu=0.5, \qquad\text{and}\qquad \sig=0.1,
\end{equation}
implying that the values of the dimensionless parameters defined
in \eqref{eq:our-gamma} and \eqref{eq:our-rho} are
$\gamma \approx 1.1$ and $\rho \approx 2.7$.

\begin{figure}[!t]
\begin{center}
\begin{tabular}{cc}
%(a) & (b) \vspace{0.5cm} \\
\includegraphics[height=2.5in,width=3in,angle=0]{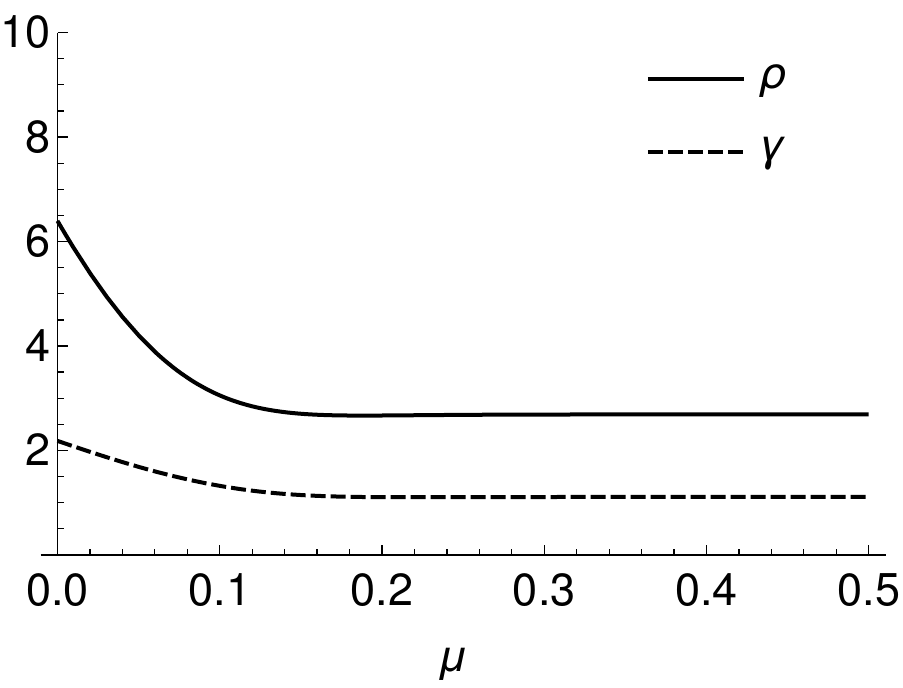} &
\includegraphics[height=2.5in,width=3in,angle=0]{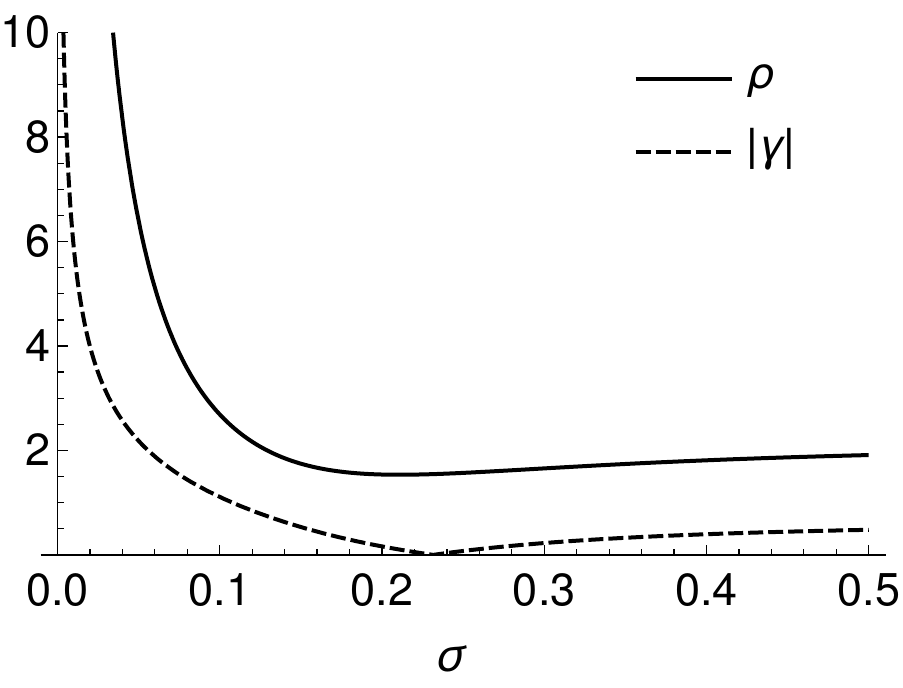}
%{\psfig{figure=neighbor1.ps,height=2.5in,width=2in,angle=-90}} &
%{\psfig{figure=neighbor2.ps,height=2.5in,width=2in,angle=-90}}
\end{tabular}
\caption{Impact of the signal location and scale parameters $\mu$ and
   $\sig$ on the quantities $\ga$ and $\rho$. In the left/right panel,
   the scale/location parameter is fixed at $\sig=0.1$/$\mu=0.5$.}
\label{fig:mu-sig-dependence}
\end{center}
\end{figure}

Figure~\ref{fig:mu-sig-dependence} gives an idea of the
impact of the signal location and scale parameters $\mu$ and $\sig$ on
$\ga$ and $\rho$. In each respective
panel, the value of the parameter not being varied is as specified in
\eqref{eq:back-signal-sets}. While the curves for
$\mu$ remain fairly flat, it is noteworthy that $\sig$ has a
dramatic effect on both quantities as it approaches the origin. This
effect is better understood when we note that for small~$\sig$,
\[ 
\rho \approx \frac{ \E_s\left[ \frac{s^3}{b^3} \right]}{
  \left(\E_s\left[ \frac{s}{b} \right]\right)^{2}} = \frac{1}{\sqrt{2
    \pi}} \sigma^{-1},\qquad\text{and}\qquad \gamma \approx \frac{ \E_s\left[ \frac{s^2}{b^2} \right]}{ \left(\E_s\left[ \frac{s}{b} \right]\right)^{3/2}} = \sqrt{\frac{2}{3}} \,\pi^{-1/4} \sigma^{-1/2}.
\]   
%%%%%%%%%%%%%%%%%%%%%%%%%%%%%%%%%%%%
%%%%%%%%%%%%%%%%%%%%%%%%%%%%%%%%%%%%

\begin{figure}[!t]
\begin{center}
\includegraphics[scale=1]{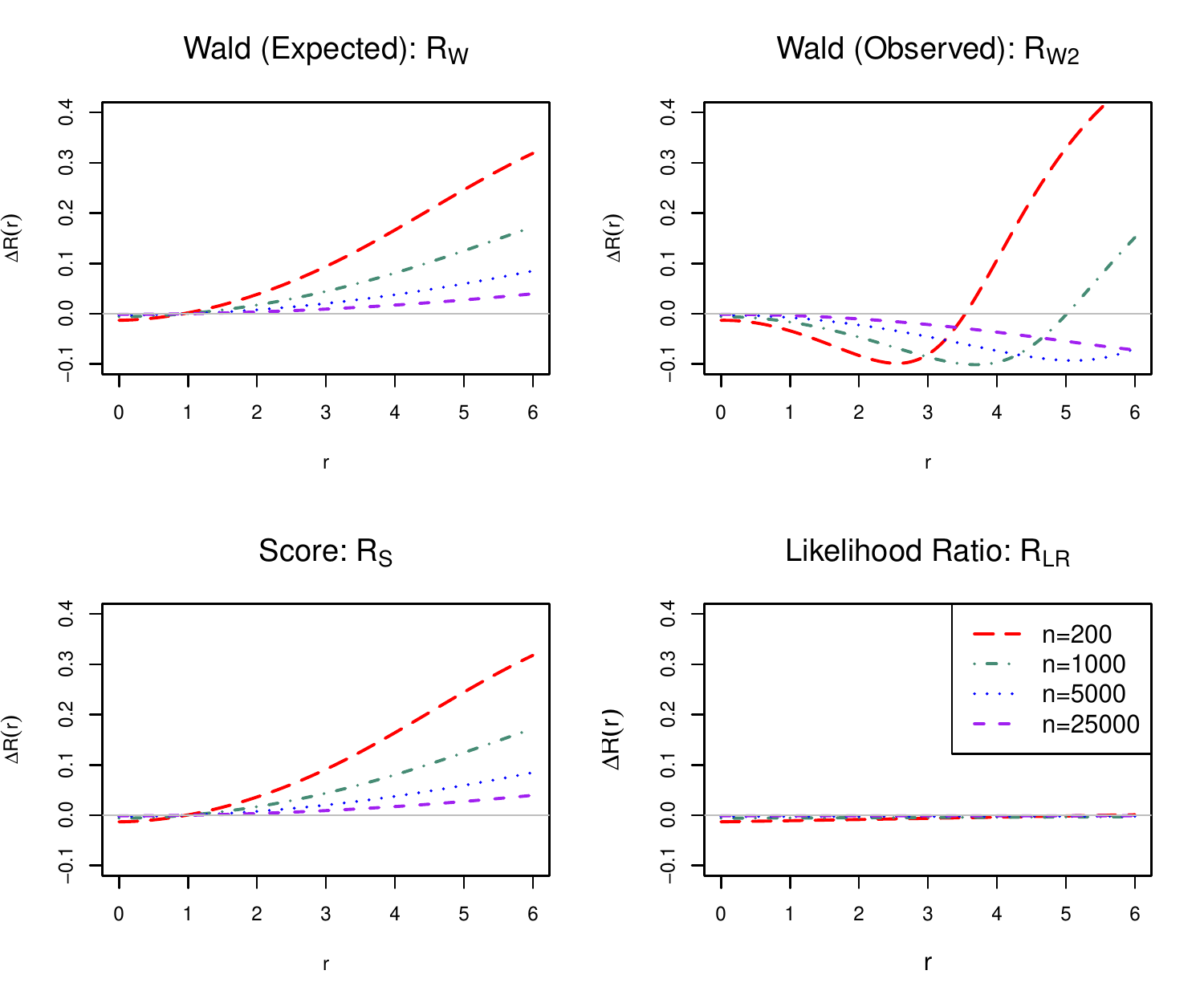}
\caption{Effective normal approximation error $\Delta R(r)$ for the signed versions
of the expected Wald ($\Rwe$), observed Wald ($\Rwe$), Score ($\Rsc$), and LR
 ($\Rlr$) statistics,  constructed from random samples of size $n$ from
model \eqref{eq:model} under signal and background densities
\eqref{eq:back-signal-model} with $\mu=0.5$ and $\sig=0.1$.}
\label{fig:Delta-R}
\end{center}
\end{figure}

To give a sense of the effective normal approximation error for some
of the statistics constructed from random samples of size $n$ from
model \eqref{eq:back-signal-model}--\eqref{eq:back-signal-sets}, Figure~\ref{fig:Delta-R} plots the value
$\Delta R(r)$ defined in \eqref{eq:Delta-R} as a function of the
observed statistic value $r$, for $\Rwe$ and
$\Rlr$. (A standard normal distribution would  have a deviation
of $\Delta R(r)=0$.) The large deviations of the un-normalized $\Rwe$
values from its third-order normal-corrected version are
particularly striking, especially at the lower sample sizes. To
understand what this means, note that the deviation of $\Rwe$ at $r=5$
for $n=200$
is approximately $\Delta\Rwe(5)=0.25$, which, for the
$Z\sim\mN$ reference distribution under $\mH_0$, translates into the
\pval{} being smaller by a factor of $P(Z>5)/P(Z>5.25)\approx 3.8$ (a
higher signal significance would be claimed than what is supported by
the data). 

In stark contrast to this is the (by comparison)
exceptionally low  normal approximation error of $\Rlr$, e.g., at $r=4$
for $n=200$, $\Delta\Rlr(4)=-0.005$, which translates into the
\pval{} being off by only a factor of $P(Z>4)/P(Z>3.995)\approx
0.98$. A glance at Table \ref{tab:cums-of-test-stats} reveals a
possible reason for the good performance of the LR statistic: the
 $O(n^{-3/2})$ values for the 3rd and 4th cumulants are zero. Since these cumulants
are precisely the ones appearing with the high order Hermite
polynomials in \eqref{eq:statexpansion}, they can contribute
substantially to the magnitude of the Edgeworth approximation at large
values of $r$. It was conjectured in~\cite{ref:mykland-1999}
that this effect may be largely responsible for the
near-optimality of $\Rlr$.

%%%%%%%%%%%%%%%%%%%%%%%%%%%%%%%%%%%%
%%%%%%%%%%%%%%%%%%%%%%%%%%%%%%%%%%%%

%%%%%%%%%%%%%%%%%%%%%%%%%%%%%%%%%%%%%%%%%%%%%%%%%%%%%%%%%%%%%%%%%%%%%%%%%%%
\section{Numerical Simulations}\label{sec:simulations}
This section undertakes  an extensive investigation of the
properties of the statistics in Table \ref{tab:cums-of-test-stats}.
As the data generating process, we take
the same flat background superimposed with 
Gaussian signal model as specified by~\eqref{eq:back-signal-model}
and \eqref{eq:back-signal-sets}. For each of the
sample sizes $n=\{200,1000,5000,25000\}$, a~total of $m=10^9$
pseudo-experiments (replications) are carried out whereby a dataset of
size $n$ is independently drawn from the model (with $\al=0$) via Monte Carlo methods,
in order to calculate the statistics in question.  

Table~\ref{tab:simulations-means} in
Appendix~\ref{app:monte-carlo-tables} compares the (Edgeworth-predicted) analytical expected value  of the
statistic (first cumulant from Table \ref{tab:cums-of-test-stats})
with its simulation-based empirical estimate (over the $m$
replications). The latter can be taken to be our best estimate of the true value;
the extent to which it deviates from truth due to Monte Carlo error being quantified by the
simulation uncertainty (in this case the standard error of the sample mean). Similar tables for the
analytical standard deviation minus one, skewness, and excess kurtosis
coefficients are given in
Tables~\ref{tab:simulations-std-dev}--\ref{tab:simulations-kurtosis}. If the statistics in question were exactly
$\mathcal{N}(0,1)$, the simulated values would all be zero (to within
simulation uncertainty). For the three classical statistics $\Rwe$,
$\Rlr$, and $\Rsc$, the results are in excellent agreement with $O(n^{-3/2})$ predictions for
    $n\geq 1000$; whereas for the remainder, values of $n\geq 25000$
    are generally needed. Agreement with predictions is markedly worse for $\Rwf$.

An alternative assessment via a chi-square 
goodness-of-fit test is carried out in Table
\ref{tab:simulations-chisq} of
Appendix~\ref{app:monte-carlo-tables}. The observed proportions of the
respective $m=10^9$ $R$ statistic values falling in each bin over the $m=10^9$
replications are compared with predicted probabilities for a $\mN$, and
second and third order Edgeworth expansions. We see substantial
disagreements with normality for all (unapproximated)
statistics and all sample sizes. The agreement of simulations with
Edgeworth predictions improves as one goes from $O(n^{-1})$ to
$O(n^{-3/2})$, and from lower to higher $n$, particularly for $\Rwe$,
$\Rlr$, and $\Rsc$.

Table~\ref{tab:simulations-tail-probs} proceeds in similar fashion, but
compares instead the Edgeworth-predicted tail probabilities for exceeding the
value of $r=5$, a traditional threshold
for signal discovery claims in high energy physics~\cite{ref:lyons-5sigma},
with the corresponding simulation-based
exceedances. Taking the case of table entries corresponding to $\Rwe$ for
example, the $O(n^{-3/2})$ prediction is computed as $P(\Rwe>5)$ using
\eqref{eq:Rwe-tail-prob}, whereas the simulated value is the empirical
proportion (say $\hat{p}$) of the $m=10^9$ $\Rwe$ values exceeding
$5$. The simulation uncertainty is the standard error of the
empirical proportion, $\sqrt{\hat{p}(1-\hat{p})/m}$. Taking twice the
simulation uncertainty as the metric (corresponding to 95\% confidence), the last column
indicates that predictions for $\Rwe$, $\Rsc$, and $\Rlr$ are
generally in
agreement with simulations (especially $\Rlr$), but the remaining statistics
show significant differences at lower $n$. Consequently, it appears
that one would need higher than third order based predictions in order to
obtain nominally correct $p$\,-values for these cases. Finally, note
that predictions for $\Rlr$ are also the closest to $\mN$ values, in
agreement with the conclusions drawn from Figure~\ref{fig:Delta-R}.

%%%%%%%%%%%%%%%%%%%%%%%%%%%%%%%%%%%%%
\begin{table}[tbh]
\caption{Comparison of Edgeworth-predicted survival probabilities at $r = 5$
  with simulation-based empirical exceedance proportions computed over 
  $10^9$ replications. The corresponding $\mN$ value is $2.87 \times
  10^{-7}$. Predictions that differ from simulated values by more than
  twice the simulation uncertainty appear in bold face.}%
\label{tab:simulations-tail-probs}
\centering
  \begin{tabular}{c|r|rrr|}
    \cline{2-5}
    & & \multicolumn{1}{c}{$O(n^{-3/2})$} &
    \multicolumn{1}{c}{Simulated} & Simulation \\
    & $n\,\ $ & \multicolumn{1}{c}{Prediction} &
    \multicolumn{1}{c}{Value} & Uncertainty \\
    \midrule
    \multicolumn{1}{|c|}{$\mbox{ }$}      & 200  &  $9.98 \times 10^{-7}$ & $10.55 \times 10^{-7} $ & $0.32 \times 10^{-7}$ \\
    \multicolumn{1}{|c|}{$\Rwe$}     & 1000 &  $5.44 \times 10^{-7}$ & $5.23 \times 10^{-7}$ & $0.23 \times 10^{-7}$ \\
    \multicolumn{1}{|c|}{$\mbox{ }$}      & 5000 &  $3.90 \times 10^{-7}$ & $4.04 \times 10^{-7}$ & $0.20 \times 10^{-7}$ \\
    \multicolumn{1}{|c|}{$\mbox{ }$}      & 25000 & $3.30 \times 10^{-7}$ & $3.19 \times 10^{-7}$ & $0.18 \times 10^{-7}$ \\
    \hline% Wald 2
    \multicolumn{1}{|c|}{$\mbox{ }$}      & 200  &  $\bm{15.0 \times 10^{-7}}$ & $0.61 \times 10^{-7}$ & $ 0.08 \times 10^{-7}$ \\
    \multicolumn{1}{|c|}{$\Rwo$}  & 1000 &  $\bm{2.83 \times 10^{-7}}$ & $0.90 \times 10^{-7}$ & $ 0.09 \times 10^{-7}$ \\
    \multicolumn{1}{|c|}{$\mbox{ }$}      & 5000 &  $1.76 \times 10^{-7}$ & $1.63 \times 10^{-7}$ & $ 0.13 \times 10^{-7}$ \\
    \multicolumn{1}{|c|}{$\mbox{ }$}      & 25000 & $2.16 \times 10^{-7}$ & $2.10 \times 10^{-7}$ & $ 0.14 \times 10^{-7}$ \\
    \hline% Wald 3
    \multicolumn{1}{|c|}{$\mbox{ }$}      & 200  &  $\bm{15.1 \times 10^{-7}}$ & $0.68 \times 10^{-7}$ & $ 0.08 \times 10^{-7}$ \\
    \multicolumn{1}{|c|}{$\Rwt$}  & 1000 &  $\bm{2.86 \times 10^{-7}}$ & $0.93 \times 10^{-7}$ & $ 0.10 \times 10^{-7}$ \\
    \multicolumn{1}{|c|}{$\mbox{ }$}      & 5000 &  $1.77 \times 10^{-7}$ & $1.63 \times 10^{-7}$ & $ 0.13 \times 10^{-7}$ \\
    \multicolumn{1}{|c|}{$\mbox{ }$}      & 25000 & $2.16 \times 10^{-7}$ & $2.10 \times 10^{-7}$ & $ 0.14 \times 10^{-7}$ \\
    \hline% Wald 4
    \multicolumn{1}{|c|}{$\mbox{ }$}      & 200  &  $\bm{9.45 \times 10^{-7}}$ & $0.71 \times 10^{-7}$ & $0.08 \times 10^{-7}$ \\
    \multicolumn{1}{|c|}{$\Rwf$}  & 1000 &  $\bm{2.21 \times 10^{-7}}$ & $1.07 \times 10^{-7}$ & $0.10 \times 10^{-7}$ \\
    \multicolumn{1}{|c|}{$\mbox{ }$}      & 5000 &  $1.85 \times 10^{-7}$ & $1.87 \times 10^{-7}$ & $0.14 \times 10^{-7}$ \\
    \multicolumn{1}{|c|}{$\mbox{ }$}      & 25000 & $2.27 \times 10^{-7}$ & $2.26 \times 10^{-7}$ & $0.15 \times 10^{-7}$ \\
    \hline% SLR
    \multicolumn{1}{|c|}{$\mbox{ }$}      & 200  &  $2.85 \times 10^{-7}$ & $2.82 \times 10^{-7}$ & $0.17 \times 10^{-7}$ \\
    \multicolumn{1}{|c|}{$\Rlr$} & 1000 &  $2.81 \times 10^{-7}$ & $2.70 \times 10^{-7}$ & $0.16 \times 10^{-7}$ \\
    \multicolumn{1}{|c|}{$\mbox{ }$}      & 5000 &  $2.83 \times 10^{-7}$ & $2.94 \times 10^{-7}$ & $0.17 \times 10^{-7}$ \\
    \multicolumn{1}{|c|}{$\mbox{ }$}      & 25000 & $2.85 \times 10^{-7}$ & $2.79 \times 10^{-7}$ & $0.17 \times 10^{-7}$ \\
    \hline% Score
    \multicolumn{1}{|c|}{$\mbox{ }$}      & 200  &  $9.90 \times 10^{-7}$ & $10.26 \times 10^{-7}$ & $0.32 \times 10^{-7}$ \\
    \multicolumn{1}{|c|}{$\Rsc$}     & 1000 &  $\bm{5.43 \times 10^{-7}}$ & $4.82 \times 10^{-7}$ & $0.22 \times 10^{-7}$ \\
    \multicolumn{1}{|c|}{$\mbox{ }$}      & 5000 &  $3.89 \times 10^{-7}$ & $3.59 \times 10^{-7}$ & $0.19 \times 10^{-7}$ \\
    \multicolumn{1}{|c|}{$\mbox{ }$}      & 25000 & $3.30 \times 10^{-7}$ & $3.18 \times 10^{-7}$ & $0.18 \times 10^{-7}$ \\
  \bottomrule
  \end{tabular}
\end{table}
%%%%%%%%%%%%%%%%%%%%%%%%%%%%%%%%%%%%%

\begin{figure}[!t]
\begin{center}
\includegraphics[scale=1]{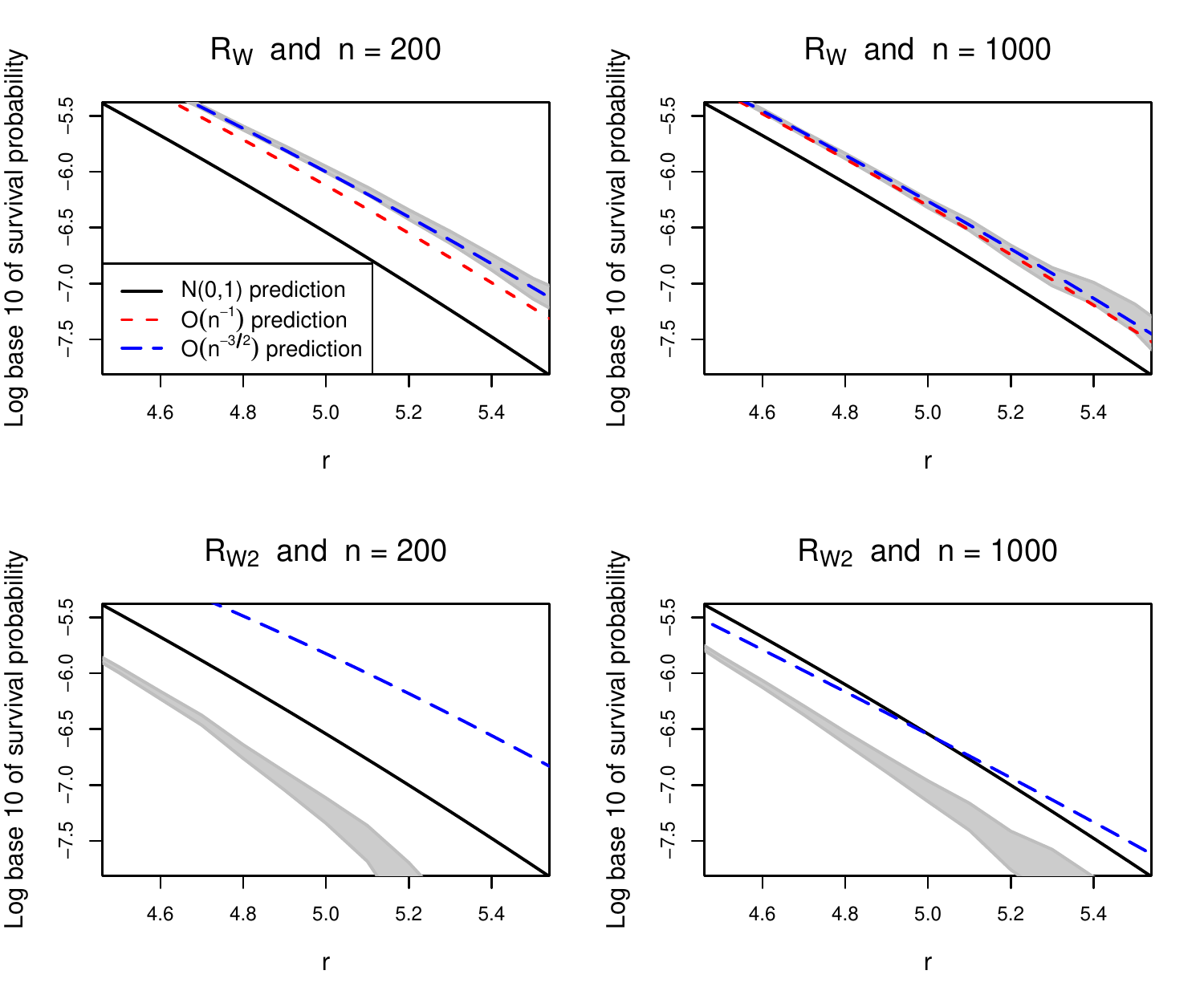}
\caption{Second and third order approximated
  Edgeworth log survival probabilities of $\Rwe$ (top panels) and $\Rwo$
  (bottom panels), computed at quantiles $r$ for $n=200$ (left panels)
  and $n=1000$ (right panels). The gray band is a 95\% simulation
  uncertainty envelope, corresponding to the true survival probability.}
\label{fig:Exceed-Wald}
\end{center}
\end{figure}

Figure~\ref{fig:Exceed-Wald} presents the survival
probabilities
% (now on log scale)
over the quantile range $4.5\leq r\leq 5.5$
relevant for establishing credibility of signal discovery claims.
The top panels show the effect of increasing the
sample size from $200$ to $1000$, as well as increasing the
order of the Edgeworth prediction from $O(n^{-1})$ to
$O(n^{-3/2})$, for $\Rwe$. Note that both predictions drift
toward the 95\% simulation uncertainty envelope (gray band) as $n$
increases; all of these converging to the $\mN$ line as expected.
An~interesting aberration occurs in the lower panels corresponding to
$\Rwo$: not only is the $O(n^{-3/2})$ prediction in the ``wrong''
direction, but the $O(n^{-1})$ one became negative
(and is thus absent from the plots). Such unexpected behavior far out in the tails, as well as the
possibility for a lower order approximation to be more accurate than a
higher order one, are documented
in the literature~\cite[sec.~5.3]{ref:severini-2000}.

%%%%%%%%%%%%%%%%%%%%%%%%%%%%%%%%%%%%
%%%%%%%%%%%%%%%%%%%%%%%%%%%%%%%%%%%%

\begin{figure}[!t]
\begin{center}
\includegraphics[scale=1]{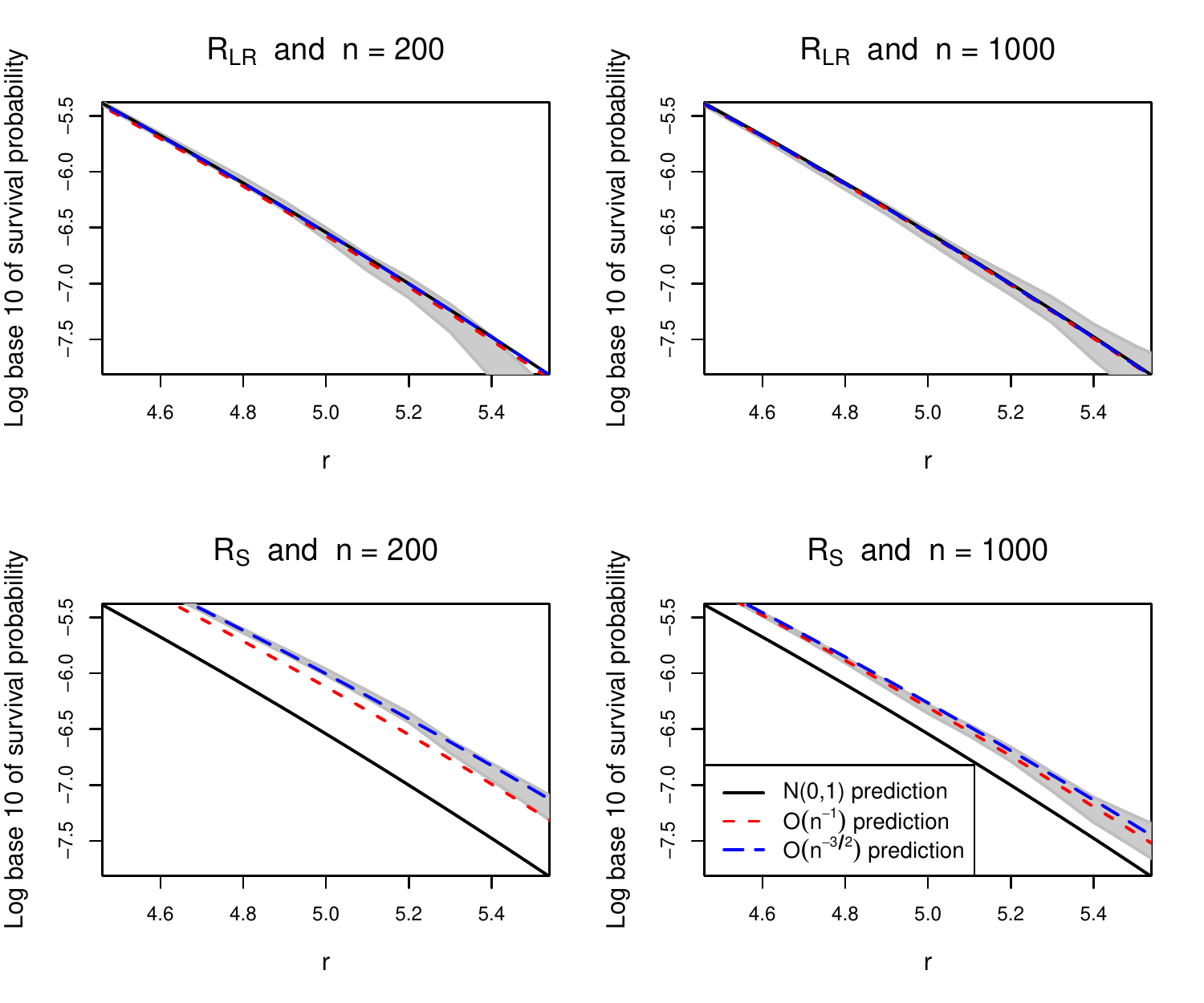}
\caption{Second and third order approximated
  Edgeworth log survival probabilities of $\Rlr$ (top panels) and $\Rsc$
  (bottom panels), computed at quantiles $r$ for $n=200$ (left panels)
  and $n=1000$ (right panels). The gray band is a 95\% simulation
  uncertainty envelope, corresponding to the true survival probability.}
\label{fig:Exceed-LRSc}
\end{center}
\end{figure}

Figure~\ref{fig:Exceed-LRSc} complements Figure~\ref{fig:Exceed-Wald}
by displaying the corresponding results for $\Rlr$ and $\Rsc$. As
noted earlier, the results for the former statistic are remarkable in
terms of: (i) close agreement with the $\mN$ line, and (ii) the relatively small
magnitude of the Edgeworth corrections. The behavior of $\Rsc$ on the
other hand is similar to that of $\Rwe$.

%%%%%%%%%%%%%%%%%%%%%%%%%%%%%%%%%%%%
%%%%%%%%%%%%%%%%%%%%%%%%%%%%%%%%%%%%

%%%%%%%%%%%%%%%%%%%%%%%%%%%%%%%%%%%%%%%%%%%%%%%%%%%%%%%%%%%%%%%%%%%%%%%%%%%
%\newpage
%\footnote{The
%  type I and II error probabilities of a hypothesis test are defined to be respectively, the
%  probability of rejecting/accepting $\mH_0$ when the truth is
%  $\mH_0$/$\mH_1$.}
\section{Study of Type I and Type II Errors}\label{sec:power}
This section undertakes a study of the type I and type II error rates for testing the null $\mH_0$ vs.~the alternative
$\mH_1$. In the context of signal
detection, these errors would correspond to false positive and false
negative probabilities, respectively.
Throughout the study we fix the nominal
(or target) probability of type I error at
$q_0 \equiv  1-\Phi(5) \approx 2.87 \times 10^{-7}$. With the model settings
once again specified by~\eqref{eq:back-signal-model} and
\eqref{eq:back-signal-sets},
we generate $m = 10^9$ replicates, each comprised of
a~random sample of size $n$ from the model.
% Thus, if $R$ is one of the statistics
% in Table \ref{tab:cums-of-test-stats} calculated from a sample of size
% $n$, we denote
To characterize the error rates, we introduce the following notation:
$\hco(n)$ is the quantile which corresponds to the predicted survival function value $q_0$.
For any statistic $R$ in Table \ref{tab:cums-of-test-stats},
this quantile is the solution of the equation $q_0=\pP(R>\hco(n))$,
where $\pP(\cdot)$ is the probability
predicted by the Edgeworth
expansions of section~\ref{subsec:strategy-one}.
Then, assuming that
$\mH_0$ is rejected if $R>\hco(n)$,
\begin{equation}\label{eq:q-one-prob}
q_1(n) \equiv \tP(R>\hco(n)|\al=0)
\end{equation}
is the probability of a type I error
(i.e., of falsely rejecting $\mH_0$ given
that the true model has $\al=0$),
estimated from simulations.
Additionally,
denote by $\tco(n)$ the quantile value established using the empirical
survival function determined via simulations. That is,
$\tco(n)$ is the solution of the equation $q_0=\tP(R>\tco(n)|\al=0)$.

Table~\ref{tab:test-stat-quantiles} gives the values of the
 predicted and  simulated,
$\hco(n)$ and $\tco(n)$, $(1-q_0)$-th quantiles. The
uncertainty here is approximated by the asymptotic standard error of the empirical $(1-q_0)$-th
quantile, given by $\sqrt{(1-q_0)q_0/[mf(\hco(n))^2]}$, where
$f(z)=-d\pS(z)/dz$. Since this density function is only
available for the predicted quantiles, we attach the uncertainty to
these values in the form of plus and minus the standard error.

%%%%%%%%%%%%%
\begin{table}[h!]
\caption{$O(n^{-3/2})$ Edgeworth predictions for the $(1-q_0)$-th  quantiles of the statistics
in Table \ref{tab:cums-of-test-stats}, compared to their corresponding values
computed based on $10^9$ simulations. The uncertainty in the predicted values appends plus and minus the
standard error for the empirical quantile. Predictions that deviate
by more than twice the uncertainty from their simulated values are bolded.}%
\label{tab:test-stat-quantiles}
\centering
\begin{tabular}{|c|l|cccc|}
\hline
 & & \multicolumn{4}{c|}{Predicted/Simulated $(1-q_0)$-th Quantile} \\
Statistic & Method  & $n=200$  & $n=1000$  & $n=5000$ & $n=25000$  \\
\midrule
\multirow{2}{*}{$\Rwe$} & Predicted & 
$5.267\pm 0.012$ & $5.131\pm 0.012$ & $5.061\pm 0.012$ & $5.028\pm 0.012$ \\
 & Simulated & 5.286 & 5.123 & 5.046 & 5.028  \\
\hline
\multirow{2}{*}{$\Rwo$} & Predicted &
$\bm{5.392\pm 0.013}$ & $\bm{4.998\pm 0.013}$ & $\bm{4.908\pm 0.011}$ & $4.946\pm 0.011 $ \\
 & Simulated & 4.736 & 4.790 & 4.872 & 4.943  \\
\hline
\multirow{2}{*}{$\Rwt$} & Predicted &
$\bm{5.394 \pm 0.013}$ & $\bm{5.000 \pm 0.013}$ & $\bm{4.908 \pm 0.011}$ & $4.946 \pm 0.011$ \\
 & Simulated & 4.744 & 4.791 & 4.872 & 4.944  \\
\hline
\multirow{2}{*}{$\Rwf$} & Predicted &
$\bm{5.292\pm 0.014}$ & $\bm{4.945\pm 0.013}$ & $\bm{4.918\pm 0.011}$ & $4.956\pm 0.011$ \\
 & Simulated & 4.755 & 4.822 & 4.893 & 4.954  \\
\hline
\multirow{2}{*}{$\Rlr$} & Predicted &
$4.999 \pm 0.011 $ &$4.996 \pm 0.011 $ &$4.998 \pm 0.011 $ &$4.999 \pm 0.011 $ \\
 & Simulated & 4.990 & 4.988 & 4.983 & 4.998  \\
\hline
\multirow{2}{*}{$\Rsc$} & Predicted &
$5.265 \pm 0.012 $&  $5.131 \pm 0.012 $&  $5.061 \pm 0.012 $ &  $5.028 \pm 0.012 $ \\
 & Simulated & 5.259 & 5.121 & 5.050 & 5.026  \\
\bottomrule
\end{tabular}
\end{table}
%%%%%%%%%%%%%

The $\hco(n)$ predictions from Table~\ref{tab:test-stat-quantiles} can
now be used as the basis for a type I and II error assessment. We start by
investigating the nominal type I error probability $q_1(n)$ as defined
in~\eqref{eq:q-one-prob}.  These results, computed
empirically based on the $m$ Monte Carlo replicates, are presented in
Table~\ref{tab:empsurv}. 
%%%%%%%%%%%%%
\begin{table}[h!]
  \caption{Type I error probabilities $q_1(n)$ as defined
in~\eqref{eq:q-one-prob}. The survival function values are obtained
based on $10^9$ simulated replicates for the predicted quantiles
$\hco(n)$ from Table~\ref{tab:test-stat-quantiles}. Values that deviate
by more than twice the simulation uncertainty of $0.17\times 10^{-7}$ 
from the nominal value of $q_0=2.87\times 10^{-7}$ are bolded.}
  \label{tab:empsurv}
  \begin{center}
    \begin{tabular}{|r|cccccc|}
      \hline
      \multicolumn{1}{|c|}{$n$}   & $\Rwe$ & $\Rwo$ & $\Rwt$ & $\Rwf$ & $\Rlr$ & $\Rsc$ \\
      \midrule
      200   & $3.03 \times 10^{-7}$ & $\bm{0.07 \times 10^{-7}}$ & $\bm{0.07 \times 10^{-7}}$ & $\bm{0.13 \times 10^{-7}}$ & $2.68 \times 10^{-7}$ & $2.73 \times 10^{-7}$ \\
      1000  & $2.71 \times 10^{-7}$ & $\bm{0.78 \times 10^{-7}}$ & $\bm{0.78 \times 10^{-7}}$ & $\bm{1.35 \times 10^{-7}}$ & $2.77 \times 10^{-7}$ & $2.72 \times 10^{-7}$ \\
      5000  & $2.65 \times 10^{-7}$ & $\bm{2.44 \times 10^{-7}}$ & $\bm{2.44 \times 10^{-7}}$ & $2.56 \times 10^{-7}$ & $2.72 \times 10^{-7}$ & $2.77 \times 10^{-7}$ \\
      25000 & $2.87 \times 10^{-7}$ & $2.82 \times 10^{-7}$ & $2.82 \times 10^{-7}$ & $2.83 \times 10^{-7}$ & $2.83 \times 10^{-7}$ & $2.86 \times 10^{-7}$ \\
      \bottomrule
    \end{tabular}
  \end{center}
\end{table}
%%%%%%%%%%%%%
If the $O(n^{-3/2})$ predictions were exact,
    we would expect approximately 95\% of all values to be within twice the simulation uncertainty of $0.17\times 10^{-7}$ 
from the nominal value of $q_0$. We once again note that the three classical statistics, $\Rwe$,
$\Rlr$, and $\Rsc$, are the only ones to perform at nominal levels (to
within Monte Carlo uncertainty) for all sample sizes. 

For the statistical power
study, we fix the probability of type I error at
$q_0$, and determine the
probability of type II error, denoted by $\err(n,\al_1)=P(R\leq\hco(n)|\al=\al_1)$.
Implicit in the notation is the fact that $\err(\cdot)$
will vary with both $n$ and the actual model value for the signal fraction
 of $\al=\al_1$ under $\mH_1$. (Also note that $q_1(n)=1-\err(n,0)$.)
For each $n$, we set the signal fraction at $\al_1=5 \sigmle$,
where $\sigmle=I(0)^{-1/2}$ is the Cramer-Rao uncertainty of $\alpha$
under $\mH_0$ (i.e., the asymptotic standard error of the
MLE for true $\alpha =0$). The values
of $\sigmle$ for the $s(x)$ and $b(x)$ used in this study
are given in Table~\ref{tab:sigmaCR}.
%%%%%%%%%%%%%%%%%%%%%%%
\begin{table}[h!]
  \caption{Values of the Cramer-Rao uncertainty for $\mH_0$,
    $\sigmle$, and corresponding values of $\al=\alpha_1$ used as the
    actual model signal fraction under $\mH_1$.}
  \label{tab:sigmaCR}
  \begin{center}
    \begin{tabular}{|r|c|c|}
      \hline
      \multicolumn{1}{|c|}{$n$}   & $\sigmle$ & $\alpha_1 = 5 \sigmle$ \\
      \midrule
      200   & $5.2401 \times 10^{-2}$ & $0.26200$ \\
      1000  & $2.3434 \times 10^{-2}$ & $0.11717$ \\
      5000  & $1.0480 \times 10^{-2}$ & $0.05240$ \\
      25000 & $0.4687 \times 10^{-2}$ & $0.02343$ \\
      \bottomrule
    \end{tabular}
  \end{center}
\end{table}
%%%%%%%%%%%%%%%%%%%%%%%

Table~\ref{tab:typeIIerrcalc} gives the resulting probability of type
II error. These are  determined empirically from the simulations
by computing the proportion of samples with $\hat{\alpha}$ below the
corresponding predicted quantiles $\hco(n)$, as well as the simulated
quantiles $\tco(n)$, listed in Table~\ref{tab:test-stat-quantiles}.
As before, there is close agreement between ``predicted'' and
``simulated'' values at all sample sizes for the three classical statistics, $\Rwe$,
$\Rlr$, and $\Rsc$. Note that the nominal $q_2(n,\al_1)$ value here is $q_2(\infty,\al_1)=0.5$, this being the value that all statistics
converge to as $n\rightarrow\infty$. In line with earlier results, it
is not surprising that $\Rlr$ yields the smallest
    predicted type
II errors at each setting of $n$.
%%%%%%%%%%%%%
\begin{table}[tbh]
\caption{Type II error probabilities for testing $\mH_1$ with the quantiles listed in Table~\ref{tab:test-stat-quantiles},
    determined empirically from $10^9$ simulations with true
    $\alpha=\al_1$ as given in Table~\ref{tab:sigmaCR}. The smallest
    predicted value at each $n$ appears in bold. The simulation uncertainty
    in these results is at most $2 \times 10^{-5}$.}%
\label{tab:typeIIerrcalc}
\centering
\begin{tabular}[h!]{|c|l|cccc|}
\toprule
 & & \multicolumn{4}{c|}{Sample Size ($n$)} \\
Statistic & Quantile  & $200$  & $1000$  & $5000$ & $25000$  \\
\midrule
\multirow{2}{*}{$\Rwe$} & Predicted & 
0.59592 & 0.55021 & 0.52433 & 0.51130 \\
 & Simulated & 0.60241 & 0.54701 & 0.51848 & 0.51143  \\
\hline
\multirow{2}{*}{$\Rwo$} & Predicted &
0.80891 & 0.63069 & 0.53250 & 0.51192  \\
 & Simulated & 0.58540 & 0.54601 & 0.51775 & 0.51074  \\
\hline
\multirow{2}{*}{$\Rwt$} & Predicted &
0.80695 & 0.63082 & 0.53257 & 0.51193  \\
 & Simulated & 0.58535 & 0.54601 & 0.51775 & 0.51074  \\
\hline
\multirow{2}{*}{$\Rwf$} & Predicted &
0.77782 & 0.59609 & 0.52830 & 0.51156  \\
 & Simulated & 0.58515 & 0.54577 & 0.51778 & 0.51074  \\
\hline
\multirow{2}{*}{$\Rlr$} & Predicted &
{\bf 0.58873} & {\bf 0.54827} & {\bf 0.52385} & {\bf 0.51119} \\
 & Simulated & 0.58533 & 0.54501 & 0.51796 & 0.51074  \\
\hline
\multirow{2}{*}{$\Rsc$} & Predicted &
0.59510 & 0.55011 & 0.52432 & 0.51129  \\
 & Simulated & 0.59296 & 0.54672 & 0.52024 & 0.51059  \\
\bottomrule
\end{tabular}
\end{table}
%%%%%%%%%%%%%

%%%%%%%%%%%%%%%%%%%%%%%%%%%%%%%%%%%%%%%%%%%%%%%%%%%%%%%%%%%%%%%%%%%%%%%%%%%
\newpage
\section{Discussion}\label{sec:discuss}
A wide gamut of near-optimal statistics
can be used in testing for the presence of a signal under a
mixture model with unknown signal fraction.
The focus of the study presented in this manuscript
was on computing Edgeworth approximations to $p$\,-values of the asymptotic
distributions of such
  statistics under the null hypothesis of no signal, so as to provide more accurate
  inferences for finite samples. Comparisons were
  made between approximations of different orders,
  highlighting striking deviations from the target standard
  normal reference distribution
  in some cases, with consequent \pval{} inflation/deflation. Finally, the performance of the corrected statistics
  was examined in terms of false positive and false negative signal detection error rates.

Given the insights gained from this study, the following
summary remarks can be offered.
 For small and moderate sample sizes, deviations from normality for
 some likelihood-based statistics used in signal searches,
    in particular the observed version of Wald and its
    variants, can be
    significant, thus altering the false discovery error rate.
    Moreover, for narrow signals these deviations are unbounded.
    The third-order approximated versions of these statistics manifest
    a~substantially improved agreement with the behavior found in simulations.  

    It is desirable to assess the magnitude of these deviations from normality.
    Their influence can be quantified by comparing the
    approximations developed in this study, to various orders of accuracy, with the asymptotic formulae,
    for all possible nuisance parameter values of the model.
    If the finite sample effects are
    substantial, proper signal strength $p$\,-values
    should then be established by direct simulations.

In comparison with the Wald and Score tests, deviations from normality are
    substantially milder for the signed LR
    statistic. Tests based on the likelihood ratio
    should therefore be preferred. The latter also have better
    capabilities of detecting actual signals (higher power), while still maintaining
    the false discovery rate at acceptably low levels. As speculated
  by~\cite{ref:mykland-1999} who proves that the $k$-th cumulant of
  $\Rlr$ vanishes to $O(n^{-k/2})$ for all $k\geq 3$, this fact
  ``\ldots would seem
  to be the main asymptotic property governing the accuracy
  behavior\ldots'' of $\Rlr$. To further elucidate this statement,
  we derive the fourth order Edgeworth-approximated tail
  probabilities for $\Rlr$ in Appendix~\ref{app:forder-lrt}.
  As predicted, the $\kappa_3$ value is $O(n^{-3/2})$ while $\kappa_4 = \kappa_5 = 0$ to $O(n^{-2})$.
%  $\Rlr$ performs better in
%  comparison with all the other statistics due to the fact that
  The
 vanishing of the highest order cumulants results in the
 suppression of high order Hermite polynomials in the Edgeworth series, $H_k(z)$ with $k > 2$,
 and it is precisely these terms that affect the tail behavior the most.
 Other statistics do not enjoy this property, containing terms
 up to $H_8(z)$ in their $O(n^{-2})$ Edgeworth expansions.

In the cases when there are no nuisance parameters in  $b(x)$, or the
practitioner is willing to treat all nuisance parameters via the
methodology of random fields (i.e., the Gross-Vitells method),
$O(n^{-3/2})$ normalized versions of the
likelihood-based statistics via
transformation \eqref{eq:norm-transform} are suggested, facilitating
standard inferences. 
%    This $O(n^{-3/2})$ normalization procedure could also be applied
%    to the random fields in the LEE determination via the Gross-Vitells method.
    Alternatively, the global significance of the signal test statistic
    could be
    adjusted conservatively, leading
    to a~subsequent (conservative) estimate of the global $p$\,-value, by proceeding as
    follows:
\begin{itemize}
\item Let $\Delta R(r(\ta))$ denote the normal approximation error
  defined for each observed (local) $r\equiv r(\ta)$ as in
  \eqref{eq:Delta-R}. Note
  that there is now an explicit dependence of $r$ (and hence $\Delta R$)
  on the nuisance parameter vector $\ta$. 
\item Perform a search over $\ta$ in
  order to locate the value of $\ta^*=\arg\max\Delta R(r(\ta))$.
  This search can utilize the same grid as the necessary search for
  $\hat{\ta} =\arg\max\ r(\ta)$. Note  that $r(\hat{\ta})$ corresponds
  to the signal with the highest local significance.
\item Calculate the global significance of the signal. This is evaluated in
  the $\mN$ approximation via the Gross-Vitells method, and is expressed in terms
  of the global $r$~\cite{ref:aslee1-2010, ref:vitells-gross-2011}.
\item Adjust the above global $r$ by subtracting $\Delta R(r(\ta^*))$. The global
  $p$\,-value is then estimated from the adjusted $r$ according to a $\mN$.
%\item Adjust the global significance of the signal, evaluated initially in
%  the $\mN$ approximation
  % via the Gross-Vitells method
%  and expressed in terms
%  of global $r$, by subtracting $\Delta R(r(\ta^*))$. The global
%  $p$\,-value is then estimated from the adjusted $r$ according to $\mN$.
\end{itemize}

\section{Acknowledgments}

I.~Volobouev thanks the Statistics Committee of the CMS Collaboration
for comments and productive discussions. This work was supported in part
by the United States Department of Energy grant DE-SC001592.

%%%%%%%%%%%%%%%%%%%%%%%%%%%%%%%%%%%%%%%%%%%%%%%%%%%%%%%%%%%%%%%%%%%%%%%%%%%
%\newpage

%%%%%%%%%%%%%%%%%%%%%%%%%%%%%%%%%%%%%%%%%%%%%%%%%%%%%%%%%%%%%%%%%%%%%%%%%%%%%%%%%%%%%%

%%%%%%%%%%%%%%%%%%%%%%%%%%%%%%%%%%%%%%%%%%%%%%%%%%%%%%%%%%%%%%%%%%%%%%%%%%%%%%%%%%%%%%
\appendix
%%%%%%%%%%%%%%%%%%%%%%%%%%%%%%%%%%%%%%%%%%%%%%%%%%%%%%%%%%%%%%%%%%%%%%%%%%%%%%%%%%%%%%
\section{Calculation of Cumulants of the Log-Likelihood Function}\label{app:cumulant-calcs}
Recalling that $\E[\cdot]$ and $\V[\cdot]$ denote, respectively,
expectation and variance under $b(x)$, and that $n\nu_{ijkl}$ represents the
$(i,j,k,l)$ joint cumulant of $(\ell_{1}(0),\ldots,\ell_{4}(0))$, we compute
initially
\[ 
n\nu_1=\E[\ell_1(0)] =
\E\left[\sum_{i=1}^n\frac{s(x_i)-b(x_i)}{b(x_i)} \right] =
\sum_{i=1}^n\int s(x)dx-\sum_{i=1}^n\int b(x)dx = n-n = 0,
\]
whence $\nu_1=0$. Similar calculations now yield:
\begin{itemize}
\item $\nu_2=\frac{1}{n}\E[\ell_1(0)^2]=\E_s[s/b]-1$.
\item $\nu_3=\frac{1}{n}\E[\ell_1(0)^3]=\E_s[s^2/b^2]-3\E_s[s/b]+2 = \ga\nu_2^{3/2}$.
\item $\nu_{11}=\frac{1}{n}\E[\ell_1(0)\ell_2(0)] = - \ga\nu_2^{3/2}$.
\item $\nu_{001}=\frac{1}{n}\E[\ell_3(0)]=2\ga\nu_2^{3/2}$.
\item
  $\nu_{101}=\frac{1}{n}\E[\ell_1(0)\ell_3(0)]=2[\E_s(s^3/b^3)-4\E_s(s^2/b^2)+6\E_s(s/b)-3]
  = 2\rho\nu_2^{2}$.
\item $\nu_4=\frac{1}{n}\E[\ell_1(0)^4]-\frac{3}{n}[\E \ell_1(0)^2]^2=(\rho-3)\nu_2^2$.
\item $\nu_{0001}=\frac{1}{n}\E[\ell_4(0)]=-6\rho\nu_2^2$.
\item $\nu_{02}=\frac{1}{n}\V[\ell_2(0)]=(\rho-1)\nu_2^2$.
\item $\nu_{21}=\frac{1}{n}\E[\ell_1(0)^2\ell_2(0)]-\frac{1}{n}\E[
  \ell_1(0)^2]\E[\ell_2(0)]=(1-\rho)\nu_2^2$.
\end{itemize}
These results use the fact that the expressions for $\gamma$ and $\rho$ defined
in \eqref{eq:our-gamma} and \eqref{eq:our-rho} become,
$\gamma=\nu_3/\nu_2^{3/2}$, and
$\rho-3=\nu_4/\nu_2^{2}$. Straightforward computations also yield the
following expressions for the information numbers under model~\eqref{eq:model}:
\[
J(\al) = \sum_{i=1}^n\frac{\left(s(x_i)-b(x_i)\right)^2}{p(x_i|\al)^2}, \qquad\text{and}\qquad
I(\al) = n\E_s\left[ \frac{\left(s(x)-b(x)\right)^2}{s(x)p(x|\al)} \right].
\]

%%%%%%%%%%%%%%%%%%%%%%%%%%%%%%%%%%%%%%%%%%%%%%%%%%%%%%%%%%%%%%%%%%%%%%%%%%%%%%%%%%%%%%
\section{Third-Order Edgeworth Expansions for $\Rwt$ and
  $\Rwf$}\label{app:torder-w3w4}
For $\Rwt$, substitution of the cumulants in Table
\ref{tab:cums-of-test-stats}  into \eqref{eq:statexpansion} gives: 
\begin{multline*}%    \label{eq:twald3}
    P(\Rwt\leq z) = \ \Phi(z) - \phi(z)\left[
    n^{-1/2} \gamma \left(- \frac{1}{2} - \frac{1}{3} H_{2}(z) \right)
    \right. \\
 \left. + n^{-1} \left( \frac{1}{36} (63 \rho - 28 \gamma^2) z + \frac{1}{12} (5 \rho + 2 \gamma^2) H_{3}(z) +  \frac{\gamma^2}{18} H_{5}(z)  \right) + O(n^{-3/2}) \right].
 \end{multline*}

For $\Rwf$, the second cumulant behavior differs from the other cases
considered in that $\kappa_2 = 1 + O(n^{-1/2})$, so that the Edgeworth
expansion in \eqref{eq:statexpansion}  must be rederived:
\begin{multline*} %\label{eq:statexpansion2}
      F(z) = \Phi(z) - \phi(z) \bigg[ \kappa_1 + \left(\frac{1}{6}
        \kappa_3 + \frac{1}{2} \kappa_1 (\kappa_2 - 1)\right) H_2(z) +
      \frac{1}{2} (\kappa_1^2 + \kappa_2 - 1) z \\
+ \left( \frac{1}{6} \kappa_1 \kappa_3 + \frac{1}{24} \kappa_4 +
  \frac{1}{8} (\kappa_2 - 1)^2\right) H_3(z) \\
+ \frac{1}{12} (\kappa_2 - 1) \kappa_3 H_4(z) + \frac{1}{72} \kappa_3^2 H_5(z) + O(n^{-3/2}) \bigg].
\end{multline*}
Substitution of the cumulants in Table
\ref{tab:cums-of-test-stats}  into this  expression then yields:
\begin{multline*}% 
P(\Rwf\leq z) = \Phi(z) - \phi(z)
    \bigg[ n^{-1/2} \gamma \left(\frac{z}{3} - \frac{1}{2} - \frac{1}{3} H_{2}(z) \right)\\
      + n^{-1} \left( \frac{\gamma^2}{3} - \rho + \frac{1}{36} (63 \rho - 22 \gamma^2) z - \frac{2}{3} \rho H_{2}(z)\right. \\
+ \left.\frac{1}{36} (15 \rho + 8 \gamma^2) H_{3}(z) -\frac{\gamma^2}{9} H_{4}(z) +  \frac{\gamma^2}{18} H_{5}(z) \right) + O(n^{-3/2}) \bigg].
\end{multline*}

%%%%%%%%%%%%%%%%%%%%%%%%%%%%%%%%%%%%%%%%%%%%%%%%%%%%%%%%%%%%%%%%%%%%%%%%%%%%%%%%%%%%%%
\section{Fourth-Order Edgeworth Expansion for $\Rlr$}\label{app:forder-lrt}
We extend \eqref{eq:statexpansion} by computing the $O(n^{-2})$
Edgeworth expansion for the normal density:
\begin{equation}
    \label{eq:statexpansionho}
    \begin{split}
      F(z) = \ & \Phi(z) - \phi(z) \bigg[ \kappa_1  + \frac{1}{2} (\kappa_1^2 + \kappa_2 - 1) z + \left(\frac{1}{6} (\kappa_1^3 + \kappa_3) + \frac{1}{2} \kappa_1 (\kappa_2 - 1)\right) H_2(z) \\
        & + \left( \frac{1}{6} \kappa_1 \kappa_3 + \frac{1}{24} \kappa_4 \right) H_3(z)
          + \left( \frac{1}{12} \kappa_3 (\kappa_1^2 + \kappa_2 - 1) + \frac{1}{24} \kappa_1 \kappa_4 + \frac{1}{120} \kappa_5 \right) H_4(z) \\
          & + \frac{1}{72} \kappa_3^2 H_5(z) + \frac{1}{144} \kappa_3 (2 \kappa_1 \kappa_3 + \kappa_4) H_6(z)
          + \frac{1}{1296} \kappa_3^3 H_8(z) + O(n^{-2}) \bigg].
    \end{split}
\end{equation}
This expansion assumes, as is the case for $\Rlr$, that $\kappa_2 = 1 + O(n^{-1})$. (Expansions for statistics in which $\kappa_2 = 1 + O(n^{-1/2})$, e.g., $\Rwf$, are more complicated.)
To compute the cumulants of $\Rlr$, introduce, similarly to $\rho$ and
$\gamma$ defined as before, the (dimensionless)
quantity 
\[
\xi = \frac{V_5}{24 (-V_2)^{5/2}} = \frac{\E_s\left[ \frac{s^4}{b^4}
  \right] - 5\E_s\left[ \frac{s^3}{b^3} \right] + 10 \E_s\left[
    \frac{s^2}{b^2} \right] - 10\E_s\left[ \frac{s}{b} \right] +
  4}{\left( \E_s\left[ \frac{s}{b} \right] - 1 \right)^{5/2}}. 
\]
In terms of these parameters, and to an accuracy of $O(n^{-2})$, we
obtain the following expressions for the
first five cumulants of $\Rlr$:
\begin{eqnarray*}
\ka_1 &=& -\frac{\gamma}{6} n^{-1/2} + \left(\frac{\gamma}{16} -
  \frac{\gamma^3}{12} + \frac{5 \gamma \rho}{16} - \frac{11
    \xi}{40}\right) n^{-3/2}, \\ 
\ka_2 &=& 1 + \frac{18 \rho - 13 \gamma^2}{36 \, n}, \qquad
\ka_3 = \left(\frac{11 \gamma \rho}{4} - \frac{251 \gamma^3}{216} -
  \frac{9 \xi}{5}\right) n^{-3/2}, 
\end{eqnarray*}
with $\ka_4 = \ka_5 = 0$.
Substitution of these  cumulants  into
\eqref{eq:statexpansionho} gives:
\begin{multline}\label{eq:tslrho}
      Pr(\Rlr\leq z) = \Phi(z) - \phi(z) \bigg\{ n^{-1/2} \left( -\frac{\gamma}{6}  \right) + n^{-1} \left( \frac{1}{12} (3 \rho - 2 \gamma^2) z \right) \\
        + n^{-3/2} \left(\frac{\gamma}{16} - \frac{\gamma^3}{12} +
          \frac{5 \gamma \rho}{16} - \frac{11 \xi}{40} + \left[\frac{5
              \gamma \rho}{12} - \frac{71 \gamma^3}{432} - \frac{3
              \xi}{10}\right] H_2(z) \right) + O(n^{-2})  \bigg\}.
\end{multline}
      For the model settings as in the illustrative example of section~\ref{sec:example}, $\xi
      \approx 5.0$. The $O(n^{-2})$ prediction agrees with the
      simulations better than the $O(n^{-3/2})$ prediction in terms of
      the mean, standard deviation, skewness, kurtosis, and $\chi^2$
      test values listed in Tables~\ref{tab:simulations-means}--\ref{tab:simulations-chisq}.

%\begin{verbatim}
%Here are the t_SLR predictions to O(N^{-2}), to be
%compared with the simulated values in the talk.
%Not sure if there is a good place for them in the
%paper, but they illustrate nicely that the O(N^{-2})
%approximation works well for SLR already for N = 200.
%
%N = 200, mean = -0.013243026
%N = 1000, mean = -0.0058619893
%N = 5000, mean = -0.0026161525
%N = 25000, mean = -0.0011694952
%
%N = 200, skewness = -0.00081575651
%N = 1000, skewness = -7.335758e-05
%N = 5000, skewness = -6.5683894e-06
%N = 25000, skewness = -5.8762154e-07
%
%For the standard deviation and the kurtosis, the O(N^{-2})
%prediction is identical to the O(N^{-3/2}) prediction.
%
%The chi^2 test value for the O(N^{-2}) approximation
%and N = 200 is 106.8/107, CL = 0.49. For higher N
%the results are basically identical with O(N^{-3/2}).
%\end{verbatim}

%%%%%%%%%%%%%%%%%%%%%%%%%%%%%%%%%%%%%%%%%%%%%%%%%%%%%%%%%%%%%%%%%%%%%%%%%%%%%%%%%%%%%%
%\newpage
\section{Assessment of Normality for the Statistics in Table
  \ref{tab:cums-of-test-stats}}\label{app:monte-carlo-tables}
The information in
Tables \ref{tab:simulations-means}--\ref{tab:simulations-kurtosis}
gauges how closely the mean, standard deviation, skewness, and kurtosis
of the statistics in Table
  \ref{tab:cums-of-test-stats} agree with the corresponding
values predicted by $\mN$ and $O(n^{-3/2})$ models. The simulated values are constructed from $m=10^9$
replications. If $r_1,\ldots,r_m$ denote the $m$ simulated values of a
particular statistic with empirical mean $\bar{r}=\sum r_i/m$,
define the empirical $k$-th central moment as $\mu_k=\sum (r_i-\bar{r})^k/m$.
The ``simulated value ($SV)$'' and ``simulation
uncertainty $(SU)$'' quantities are, respectively, the appropriate
empirical moment estimate, and the standard deviation of the estimate, determined as
follows (see~\cite{ref:joanes-gill-1998} for details).
%, assumed to be independently drawn from a $\mN$\begin{itemize}
\begin{itemize}
\item Table \ref{tab:simulations-means} for the mean: $SV = \bar{r}$,
  and $SU = \frac{1}{\sqrt{m}} =3.2 \times 10^{-5}$.
\item Table \ref{tab:simulations-std-dev} for the standard deviation
  minus $1$ ($SU$ assumes the sample is drawn from a $\mN$): 
$SV = \sqrt{\mu_{2}}-1$, and $SU = 1/\sqrt{2(m-1)} =2.2 \times 10^{-5}$.
\item Table \ref{tab:simulations-skewness} for the skewness coefficient ($SU$  assumes the sample is drawn from a $\mN$):
\[ 
SV = \frac{\mu_3}{\mu_2^{3/2}}, \qquad SU = \sqrt{\frac{6m(m-1)}{(m-2)(m+1)(m+3)}} =7.7 \times 10^{-5}.
\]
\item Table \ref{tab:simulations-kurtosis} for the kurtosis coefficient ($SU$ assumes the sample is drawn from a $\mN$): 
\[ 
SV = \frac{\mu_4}{\mu_2^{2}}-3, \qquad SU =   \sqrt{\frac{24m(m - 1)^2}{(m - 3) (m - 2) (3 + m) (5 + m)}} =1.5 \times 10^{-4}.
\]
\end{itemize}

Table
\ref{tab:simulations-chisq} gives the results of a chi-square ($\chi^2$)
goodness-of-fit test. Using a bin width of $0.1$ (and utilizing
all bins with 25 or more predicted counts), the proportions of the respective
$R$ statistic values falling in each bin over the $m=10^9$
replications are converted to a single $\chi^2$-value comparing the
observed proportions with predicted probabilities for a $\mN$, and
second and third order Edgeworth approximations. Each $\chi^2$-value is then
converted to a corresponding \pval{} for the test by calculating the
survival probability at the value under a $\chi^2$ distribution with the indicated degrees of
freedom.

%%%%%%%%%%%%%%%%%%%%%%%%%%%%%%%%%%%%%
%%% Table of Means
%%%%%%%%%%%%%%%%%%%%%%%%%%%%%%%%%%%%%
\begin{table}[tbh]
\caption{Comparison of the first cumulant from 
  Table~\ref{tab:cums-of-test-stats} with the simulation-based empirical
  estimate constructed from $10^9$ replications.
  Predictions that differ from simulated values by
  more than twice the simulation uncertainty appear in bold face.
  The corresponding $\mN$ value is~$0$.}%
\label{tab:simulations-means}
\centering
\begin{tabular}{c|r|rrc|}
\cline{2-5}
    & & \multicolumn{1}{c}{$O(n^{-3/2})$} &
    \multicolumn{1}{c}{Simulated} & Simulation  \\
    & $n\,\ $ & \multicolumn{1}{c}{Prediction} &
    \multicolumn{1}{c}{Value} & Uncertainty \\
    \midrule
    \multicolumn{1}{|c|}{$\mbox{ }$}      & 200  &  \multicolumn{1}{c}{\bf{0}} & $-9.0 \times 10^{-5}$  & $\mbox{ }$ \\
    \multicolumn{1}{|c|}{$\Rwe$}     & 1000 &  \multicolumn{1}{c}{0} & $-0.9 \times 10^{-5}$  & $3.2 \times 10^{-5}$ \\
    \multicolumn{1}{|c|}{$\mbox{ }$}      & 5000 &  \multicolumn{1}{c}{0} & $-1.6 \times 10^{-5}$  & $\mbox{ }$ \\
    \multicolumn{1}{|c|}{$\mbox{ }$}      & 25000 & \multicolumn{1}{c}{0} & $4.8  \times 10^{-5}$  & $\mbox{ }$ \\
    \hline
    \multicolumn{1}{|c|}{$\mbox{ }$}      & 200  &  $\bf{-39.222 \times 10^{-3}}$ & $-40.724 \times 10^{-3}$ & $\mbox{ }$ \\
    \multicolumn{1}{|c|}{$\Rwo$}  & 1000 &  $\bf{-17.541 \times 10^{-3}}$ & $-17.670 \times 10^{-3}$ & $0.032 \times 10^{-3}$ \\
    \multicolumn{1}{|c|}{$\mbox{ }$}      & 5000 &  $-7.844 \times 10^{-3}$  & $-7.872  \times 10^{-3}$ & $\mbox{ }$ \\
    \multicolumn{1}{|c|}{$\mbox{ }$}      & 25000 & $-3.508 \times 10^{-3}$  & $-3.461  \times 10^{-3}$ & $\mbox{ }$ \\
    \hline
    \multicolumn{1}{|c|}{$\mbox{ }$}      & 200  &  $\bf{-39.222 \times 10^{-3}}$ & $-40.817 \times 10^{-3}$ & $\mbox{ }$ \\
    \multicolumn{1}{|c|}{$\Rwt$}  & 1000 &  $\bf{-17.541 \times 10^{-3}}$ & $-17.678 \times 10^{-3}$ & $0.032 \times 10^{-3}$ \\
    \multicolumn{1}{|c|}{$\mbox{ }$}      & 5000 &  $-7.844 \times 10^{-3}$  & $-7.872  \times 10^{-3}$ & $\mbox{ }$ \\
    \multicolumn{1}{|c|}{$\mbox{ }$}      & 25000 & $-3.508 \times 10^{-3}$  & $-3.461  \times 10^{-3}$ & $\mbox{ }$ \\
    \hline
    \multicolumn{1}{|c|}{$\mbox{ }$}      & 200  &  $\bf{-50.617 \times 10^{-3}}$ & $-48.508 \times 10^{-3}$ & $\mbox{ }$ \\
    \multicolumn{1}{|c|}{$\Rwf$}  & 1000 &  $\bf{-19.820 \times 10^{-3}}$ & $-19.106 \times 10^{-3}$ & $0.032 \times 10^{-3}$ \\
    \multicolumn{1}{|c|}{$\mbox{ }$}      & 5000 &  $\bf{-8.300 \times 10^{-3}}$  & $-8.152  \times 10^{-3}$ & $\mbox{ }$ \\
    \multicolumn{1}{|c|}{$\mbox{ }$}      & 25000 & $\bf{-3.599 \times 10^{-3}}$  & $-3.516 \times 10^{-3}$ & $\mbox{ }$ \\
    \hline
    \multicolumn{1}{|c|}{$\mbox{ }$}      & 200  &  $\bf{-13.074 \times 10^{-3}}$ & $-13.199 \times 10^{-3}$ & $\mbox{ }$ \\
    \multicolumn{1}{|c|}{$\Rlr$} & 1000 &  $-5.847 \times 10^{-3}$ & $-5.860 \times 10^{-3}$ & $0.032 \times 10^{-3}$ \\
    \multicolumn{1}{|c|}{$\mbox{ }$}      & 5000 &  $-2.615 \times 10^{-3}$ & $-2.631 \times 10^{-3}$ & $\mbox{ }$ \\
    \multicolumn{1}{|c|}{$\mbox{ }$}      & 25000 & $-1.169 \times 10^{-3}$ & $-1.121 \times 10^{-3}$ & $\mbox{ }$ \\
    \hline
    \multicolumn{1}{|c|}{$\mbox{ }$}      & 200  &  \multicolumn{1}{c}{0} & $ 2.7 \times 10^{-5}$  & $\mbox{ }$ \\
    \multicolumn{1}{|c|}{$\Rsc$}     & 1000 &  \multicolumn{1}{c}{0} & $-1.6 \times 10^{-5}$  & $3.2 \times 10^{-5}$ \\
    \multicolumn{1}{|c|}{$\mbox{ }$}      & 5000 &  \multicolumn{1}{c}{0} & $1.3 \times 10^{-5}$  & $\mbox{ }$ \\
    \multicolumn{1}{|c|}{$\mbox{ }$}      & 25000 & \multicolumn{1}{c}{0} & $0.1 \times 10^{-5}$  & $\mbox{ }$ \\
\bottomrule
\end{tabular}
\end{table}
%%%%%%%%%%%%%%%%%%%%%%%%%%%%%%%%%%%%%

%%%%%%%%%%%%%%%%%%%%%%%%%%%%%%%%%%%%%
%%% Table of Std devs minus one
%%%%%%%%%%%%%%%%%%%%%%%%%%%%%%%%%%%%%
\begin{table}[tbh]
\caption{Comparison of the standard deviation minus one, constructed from the
  cumulants in Table~\ref{tab:cums-of-test-stats}, with
  the simulation-based empirical estimate over $10^9$ replications.
  Predictions that differ from simulated values by
  more than twice the simulation uncertainty appear in bold face.  
  The corresponding $\mN$ value is~$0$.}%
\label{tab:simulations-std-dev}
\centering
  \begin{tabular}{c|r|rrc|}
    \cline{2-5}
    & & \multicolumn{1}{c}{$O(n^{-3/2})$} & \multicolumn{1}{c}{Simulated} & Simulation \\
    & $n\,\ $ & \multicolumn{1}{c}{Prediction} & \multicolumn{1}{c}{Value} & Uncertainty \\
    \midrule
    \multicolumn{1}{|c|}{$\mbox{ }$}      & 200  &  $\bf{11.46 \times 10^{-4}}$ & $11.97 \times 10^{-4}$ & $\mbox{ }$ \\
    \multicolumn{1}{|c|}{$\Rwe$}     & 1000 &  $2.29 \times 10^{-4}$ & $2.60 \times 10^{-4}$ & $0.22 \times 10^{-4}$ \\
    \multicolumn{1}{|c|}{$\mbox{ }$}      & 5000 &  $0.46 \times 10^{-4}$ & $0.76 \times 10^{-4}$ & $\mbox{ }$ \\
    \multicolumn{1}{|c|}{$\mbox{ }$}      & 25000 & $0.09 \times 10^{-4}$ & $-0.30 \times 10^{-4}$ & $\mbox{ }$ \\
    \hline
    \multicolumn{1}{|c|}{$\mbox{ }$}      & 200  &  $\bf{161.93 \times 10^{-4}}$ & $174.00 \times 10^{-4}$ & $\mbox{ }$ \\
    \multicolumn{1}{|c|}{$\Rwo$}  & 1000 &  $\bf{32.59 \times 10^{-4}}$ & $33.31 \times 10^{-4}$ & $0.22 \times 10^{-4}$ \\
    \multicolumn{1}{|c|}{$\mbox{ }$}      & 5000 &  $6.53 \times 10^{-4}$ & $6.86 \times 10^{-4}$ & $\mbox{ }$ \\
    \multicolumn{1}{|c|}{$\mbox{ }$}      & 25000 & $1.31 \times 10^{-4}$ & $0.92 \times 10^{-4}$ & $\mbox{ }$ \\
    \hline
    \multicolumn{1}{|c|}{$\mbox{ }$}      & 200  &  $\bf{178.17 \times 10^{-4}}$ & $191.06 \times 10^{-4}$ & $\mbox{ }$ \\
    \multicolumn{1}{|c|}{$\Rwt$}  & 1000 &  $\bf{35.89 \times 10^{-4}}$  & $36.63 \times 10^{-4}$ & $0.22 \times 10^{-4}$ \\
    \multicolumn{1}{|c|}{$\mbox{ }$}      & 5000 &  $7.19 \times 10^{-4}$   & $7.52 \times 10^{-4}$ & $\mbox{ }$ \\
    \multicolumn{1}{|c|}{$\mbox{ }$}      & 25000 & $1.44 \times 10^{-4}$   & $1.05 \times 10^{-4}$ & $\mbox{ }$ \\
    \hline
    \multicolumn{1}{|c|}{$\mbox{ }$}      & 200  &  $\bf{441.74 \times 10^{-4}}$ & $491.83 \times 10^{-4}$ & $\mbox{ }$ \\
    \multicolumn{1}{|c|}{$\Rwf$}  & 1000 &  $\bf{153.76 \times 10^{-4}}$ & $157.40 \times 10^{-4}$ & $0.22 \times 10^{-4}$ \\
    \multicolumn{1}{|c|}{$\mbox{ }$}      & 5000 &  $\bf{59.72 \times 10^{-4}}$  & $60.30 \times 10^{-4}$ & $\mbox{ }$ \\
    \multicolumn{1}{|c|}{$\mbox{ }$}      & 25000 & $24.88 \times 10^{-4}$  & $24.51 \times 10^{-4}$ & $\mbox{ }$ \\
    \hline
    \multicolumn{1}{|c|}{$\mbox{ }$}      & 200  &  $22.48 \times 10^{-4}$ & $22.86 \times 10^{-4}$ & $\mbox{ }$ \\
    \multicolumn{1}{|c|}{$\Rlr$} & 1000 &  $4.50 \times 10^{-4}$  & $4.79 \times 10^{-4}$ & $0.22 \times 10^{-4}$ \\
    \multicolumn{1}{|c|}{$\mbox{ }$}      & 5000 &  $0.90 \times 10^{-4}$  & $1.21 \times 10^{-4}$ & $\mbox{ }$ \\
    \multicolumn{1}{|c|}{$\mbox{ }$}      & 25000 & $0.18 \times 10^{-4}$  & $-0.21 \times 10^{-4}$ & $\mbox{ }$ \\
    \hline
    \multicolumn{1}{|c|}{$\mbox{ }$}      & 200  &  \multicolumn{1}{c}{0} & $-1.5 \times 10^{-5}$  & $\mbox{ }$ \\
    \multicolumn{1}{|c|}{$\Rsc$}     & 1000 &  \multicolumn{1}{c}{0} & $0.1 \times 10^{-5}$  & $2.2 \times 10^{-5}$ \\
    \multicolumn{1}{|c|}{$\mbox{ }$}      & 5000 &  \multicolumn{1}{c}{0} & $3.8 \times 10^{-5}$  & $\mbox{ }$ \\
    \multicolumn{1}{|c|}{$\mbox{ }$}      & 25000 & \multicolumn{1}{c}{\bf{0}} & $4.7 \times 10^{-5}$  & $\mbox{ }$ \\
\bottomrule
  \end{tabular}
\end{table}

%%%%%%%%%%%%%%%%%%%%%%%%%%%%%%%%%%%%%
%%% Table of skewness
%%%%%%%%%%%%%%%%%%%%%%%%%%%%%%%%%%%%%
\begin{table}[tbh]
\caption{Comparison of the skewness coefficient, constructed from  the
  cumulants in Table~\ref{tab:cums-of-test-stats}, with the simulation-based empirical estimate over $10^9$ replications.
  Predictions that differ from simulated values by
  more than twice the simulation uncertainty appear in bold face.
  The corresponding $\mN$ value is~$0$.}%
\label{tab:simulations-skewness}
\centering
  \begin{tabular}{c|r|rrc|}
    \cline{2-5}
    & & \multicolumn{1}{c}{$O(n^{-3/2})$} & \multicolumn{1}{c}{Simulated} & Simulation \\
    & $n\,\ $ & \multicolumn{1}{c}{Prediction} & \multicolumn{1}{c}{Value} & Uncertainty \\
    \midrule
    \multicolumn{1}{|c|}{$\mbox{ }$}      & 200  &  \bf{0.07817} & 0.07747 &  \\
    \multicolumn{1}{|c|}{$\Rwe$}     & 1000 &  0.03506 & 0.03513 & 0.00008 \\
    \multicolumn{1}{|c|}{$\mbox{ }$}      & 5000 &  0.01569 & 0.01562 &  \\
    \multicolumn{1}{|c|}{$\mbox{ }$}      & 25000 & 0.00702 & 0.00701 &  \\
    \hline
    \multicolumn{1}{|c|}{$\mbox{ }$}      & 200  &  $\bf{-0.14951}$ & $-0.17543$ &  \\
    \multicolumn{1}{|c|}{$\Rwo$}  & 1000 &  $\bf{-0.06948}$ & $-0.07150$ & 0.00008 \\
    \multicolumn{1}{|c|}{$\mbox{ }$}      & 5000 &  $\bf{-0.03132}$ & $-0.03157$ &  \\
    \multicolumn{1}{|c|}{$\mbox{ }$}      & 25000 & $-0.01402$ & $-0.01405$ &  \\
    \hline
    \multicolumn{1}{|c|}{$\mbox{ }$}      & 200  &  $\bf{-0.14879}$ & $-0.17563$ &  \\
    \multicolumn{1}{|c|}{$\Rwt$}  & 1000 &  $\bf{-0.06941}$ & $-0.07152$ & 0.00008 \\
    \multicolumn{1}{|c|}{$\mbox{ }$}      & 5000 &  $\bf{-0.0313}1$ & $-0.03157$ &  \\
    \multicolumn{1}{|c|}{$\mbox{ }$}      & 25000 & $-0.01402$ & $-0.01405$ &  \\
    \hline
    \multicolumn{1}{|c|}{$\mbox{ }$}      & 200  &  $\bf{-0.17965}$ & $-0.21724$ &  \\
    \multicolumn{1}{|c|}{$\Rwf$}  & 1000 &  $\bf{-0.07612}$ & $-0.07883$ & 0.00008 \\
    \multicolumn{1}{|c|}{$\mbox{ }$}      & 5000 &  $\bf{-0.03269}$ & $-0.03300$ &  \\
    \multicolumn{1}{|c|}{$\mbox{ }$}      & 25000 & $-0.01431$ & $-0.01434$ &  \\
    \hline
    \multicolumn{1}{|c|}{$\mbox{ }$}      & 200  &  \multicolumn{1}{c}{\bf{0}} & $-0.00084$ &  \\
    \multicolumn{1}{|c|}{$\Rlr$} & 1000 &  \multicolumn{1}{c}{0} & $0.00006$ & 0.00008 \\
    \multicolumn{1}{|c|}{$\mbox{ }$}      & 5000 &  \multicolumn{1}{c}{0} & $-0.00007$  &  \\
    \multicolumn{1}{|c|}{$\mbox{ }$}      & 25000 & \multicolumn{1}{c}{0} & $-0.00001$  &  \\
    \hline
    \multicolumn{1}{|c|}{$\mbox{ }$}      & 200  &  0.07844 & 0.07853 &  \\
    \multicolumn{1}{|c|}{$\Rsc$}     & 1000 &  0.03508 & 0.03501 & 0.00008 \\
    \multicolumn{1}{|c|}{$\mbox{ }$}      & 5000 &  0.01569 & 0.01567 &  \\
    \multicolumn{1}{|c|}{$\mbox{ }$}      & 25000 & 0.00702 & 0.00693 &  \\
  \bottomrule
  \end{tabular}
\end{table}

%%%%%%%%%%%%%%%%%%%%%%%%%%%%%%%%%%%%%
%%% Table of excess kurtosis
%%%%%%%%%%%%%%%%%%%%%%%%%%%%%%%%%%%%%
\begin{table}[tbh]
\caption{Comparison of the excess kurtosis coefficient, constructed  from  the
  cumulants in  Table~\ref{tab:cums-of-test-stats}, with the simulation-based empirical estimate over $10^9$ replications.
  Predictions that differ from simulated values by
  more than twice the simulation uncertainty appear in bold face.
  The corresponding $\mN$ value is~$0$.}%
\label{tab:simulations-kurtosis}
\centering
  \begin{tabular}{c|r|rrc|}
    \cline{2-5}
    & & \multicolumn{1}{c}{$O(n^{-3/2})$} & \multicolumn{1}{c}{Simulated} & Simulation \\
    & $n\,\ $ & \multicolumn{1}{c}{Prediction} & \multicolumn{1}{c}{value} & Uncertainty \\
    \midrule
    \multicolumn{1}{|c|}{$\mbox{ }$}      & 200  &  $\bf{-15.5 \times 10^{-4}}$ & $-10.6 \times 10^{-4} $ & \\
    \multicolumn{1}{|c|}{$\Rwe$}     & 1000 &  $-3.1 \times 10^{-4}$ & $-3.6 \times 10^{-4}$ & $1.5 \times 10^{-4}$ \\
    \multicolumn{1}{|c|}{$\mbox{ }$}      & 5000 &  $-0.6 \times 10^{-4}$ & $-2.7 \times 10^{-4}$ & \\
    \multicolumn{1}{|c|}{$\mbox{ }$}      & 25000 & $-0.1 \times 10^{-4}$ & $2.6 \times 10^{-4}$ & \\
    \hline
    \multicolumn{1}{|c|}{$\mbox{ }$}      & 200  &  $\bf{126.10 \times 10^{-3}}$ & $163.24 \times 10^{-3}$ & \\
    \multicolumn{1}{|c|}{$\Rwo$}  & 1000 &  $\bf{26.54 \times 10^{-3}}$  & $27.76 \times 10^{-3}$ & $0.15 \times 10^{-3}$ \\
    \multicolumn{1}{|c|}{$\mbox{ }$}      & 5000 &  $5.36 \times 10^{-3}$   & $5.24 \times 10^{-3}$ & \\
    \multicolumn{1}{|c|}{$\mbox{ }$}      & 25000 & $1.08 \times 10^{-3}$   & $1.35 \times 10^{-3}$ & \\
    \hline
    \multicolumn{1}{|c|}{$\mbox{ }$}      & 200  &  $\bf{125.29 \times 10^{-3}}$ & $163.60 \times 10^{-3}$ & \\
    \multicolumn{1}{|c|}{$\Rwt$}  & 1000 &  $\bf{26.51 \times 10^{-3}}$  & $ 27.77 \times 10^{-3}$ & $0.15 \times 10^{-3}$ \\
    \multicolumn{1}{|c|}{$\mbox{ }$}      & 5000 &  $5.36 \times 10^{-3}$   & $ 5.24 \times 10^{-3}$ & \\
    \multicolumn{1}{|c|}{$\mbox{ }$}      & 25000 & $1.08 \times 10^{-3}$   & $ 1.35 \times 10^{-3}$ & \\
    \hline
    \multicolumn{1}{|c|}{$\mbox{ }$}      & 200  &  $\bf{113.11 \times 10^{-3}}$ & $207.78 \times 10^{-3}$ & \\
    \multicolumn{1}{|c|}{$\Rwf$}  & 1000 &  $\bf{25.30 \times 10^{-3}}$  & $ 30.56 \times 10^{-3}$ & $0.15 \times 10^{-3}$ \\
    \multicolumn{1}{|c|}{$\mbox{ }$}      & 5000 &  $5.25 \times 10^{-3}$   & $ 5.47 \times 10^{-3}$ & \\
    \multicolumn{1}{|c|}{$\mbox{ }$}      & 25000 & $1.07 \times 10^{-3}$   & $ 1.37 \times 10^{-3}$ & \\
    \hline
    \multicolumn{1}{|c|}{$\mbox{ }$}      & 200  &  \multicolumn{1}{c}{\bf{0}} & $3.7 \times 10^{-4}$ & \\
    \multicolumn{1}{|c|}{$\Rlr$} & 1000 &  \multicolumn{1}{c}{0} & $-0.6 \times 10^{-4}$ & $1.5 \times 10^{-4}$ \\
    \multicolumn{1}{|c|}{$\mbox{ }$}      & 5000 &  \multicolumn{1}{c}{0} & $-2.0 \times 10^{-4}$ &  \\
    \multicolumn{1}{|c|}{$\mbox{ }$}      & 25000 & \multicolumn{1}{c}{0} & $2.7 \times 10^{-4}$ &  \\
    \hline
    \multicolumn{1}{|c|}{$\mbox{ }$}      & 200  &  $-15.5 \times 10^{-4}$ & $-14.6 \times 10^{-4} $ & \\
    \multicolumn{1}{|c|}{$\Rsc$}     & 1000 &  $-3.1 \times 10^{-4}$  & $-2.7 \times 10^{-4}$ & $1.5 \times 10^{-4}$ \\
    \multicolumn{1}{|c|}{$\mbox{ }$}      & 5000 &  $-0.6 \times 10^{-4}$  & $0.7 \times 10^{-4}$ & \\
    \multicolumn{1}{|c|}{$\mbox{ }$}      & 25000 & $-0.1 \times 10^{-4}$  & $2.3 \times 10^{-4}$ & \\
  \bottomrule
  \end{tabular}
\end{table}

%%%%%%%%%%%%%%%%%%%%%%%%%%%%%%%%%%%%%
%%% Table of chisq GOF test
%%%%%%%%%%%%%%%%%%%%%%%%%%%%%%%%%%%%%
\begin{table}[tbh]
\caption{$\chi^2$ goodness-of-fit test for the statistics in Table
  \ref{tab:cums-of-test-stats} with $\mN$, and second and third order
  Edgeworth predictions. For each $n$, the test is constructed from $10^9$
replications using a bin width of $0.1$ and all bins with
25 or more predicted counts. Entries are formated as $\{a/b,\
c\}$, where $a$ is the $\chi^2$ statistic value, $b$ is the number of degrees of freedom (i.e., the number of bins used),
and $c$ is the corresponding \pval{}.}
\label{tab:simulations-chisq}
  \centering
  \begin{tabular}{c|r|r|r|r|}
    \cline{2-5}
    & &  \multicolumn{1}{c|}{$\mN$} & \multicolumn{1}{c|}{$O(n^{-1})$} & \multicolumn{1}{c|}{$O(n^{-3/2})$} \\
    & $n\,\ $ & \multicolumn{1}{c|}{prediction}      & \multicolumn{1}{c|}{prediction}              & \multicolumn{1}{c|}{prediction}            \\
    \midrule
    \multicolumn{1}{|c|}{$\mbox{ }$}      & 200  & 1012631/107, \ 0.00 & 12920/100, \ 0.00 & 358.7/107, \ 0.00 \\
    \multicolumn{1}{|c|}{$\Rwe$}     & 1000 & 205959/107, \ 0.00 & 534.4/106, \ 0.00 & 124.6/106, \ 0.10 \\
    \multicolumn{1}{|c|}{$\mbox{ }$}      & 5000 & 40702/107, \ 0.00 & 110.2/107, \ 0.40 & 90.29/107, \ 0.88 \\
    \multicolumn{1}{|c|}{$\mbox{ }$}      & 25000 & 8254/107, \ 0.00 & 81.34/107, \ 0.97 & 82.37/107, \ 0.96 \\
    \hline                                                               
    \multicolumn{1}{|c|}{$\mbox{ }$}      & 200  & 11789846/107, \ 0.00 & 3710820/92, \ 0.00 & 402208/116, \ 0.00 \\
    \multicolumn{1}{|c|}{$\Rwo$}  & 1000 & 1266791/107, \ 0.00 & 105137/101, \ 0.00 & 4104/110, \ 0.00 \\
    \multicolumn{1}{|c|}{$\mbox{ }$}      & 5000 & 231468/107, \ 0.00 & 2967/106, \ 0.00 & 169.5/107, \ 0.00 \\
    \multicolumn{1}{|c|}{$\mbox{ }$}      & 25000 & 44997/107, \ 0.00 & 196.9/107, \ 0.00 & 78.53/107, \ 0.98 \\
    \hline                                                               
    \multicolumn{1}{|c|}{$\mbox{ }$}      & 200  & 12178306/107, \ 0.00 & 3961404/92, \ 0.00 & 426251/116, \ 0.00 \\
    \multicolumn{1}{|c|}{$\Rwt$}  & 1000 & 1275312/107, \ 0.00 & 110857/101, \ 0.00 & 4292/110, \ 0.00 \\
    \multicolumn{1}{|c|}{$\mbox{ }$}      & 5000 & 231797/107, \ 0.00 & 3169/106, \ 0.00 & 171.2/107, \ 0.00 \\
    \multicolumn{1}{|c|}{$\mbox{ }$}      & 25000 & 45006/107, \ 0.00 & 201.8/107, \ 0.00 & 77.90/107, \ 0.98 \\
    \hline                                                               
    \multicolumn{1}{|c|}{$\mbox{ }$}      & 200  & 31650126/107, \ 0.00 & 5860538/95, \ 0.00 & 1352453/116, \ 0.00 \\
    \multicolumn{1}{|c|}{$\Rwf$}  & 1000 & 2166332/107, \ 0.00 & 116183/103, \ 0.00 & 8056/110, \ 0.00 \\
    \multicolumn{1}{|c|}{$\mbox{ }$}      & 5000 & 333907/107, \ 0.00 & 4365/106, \ 0.00 & 212.1/108, \ 0.00 \\
    \multicolumn{1}{|c|}{$\mbox{ }$}      & 25000 & 59517/107, \ 0.00 & 248.1/107, \ 0.00 & 87.67/107, \ 0.92 \\
    \hline                                                               
    \multicolumn{1}{|c|}{$\mbox{ }$}      & 200  & 185597/107, \ 0.00 & 11493/107, \ 0.00 & 292.0/107, \ 0.00 \\
    \multicolumn{1}{|c|}{$\Rlr$} & 1000 & 34917/107, \ 0.00 & 616.9/107, \ 0.00 & 126.5/107, \ 0.10 \\
    \multicolumn{1}{|c|}{$\mbox{ }$}      & 5000 & 7068/107, \ 0.00 & 142.5/107, \ 0.01 & 113.8/107, \ 0.31 \\
    \multicolumn{1}{|c|}{$\mbox{ }$}      & 25000 & 1339/107, \ 0.00 & 80.78/107, \ 0.97 & 82.95/107, \ 0.96 \\
    \hline
    \multicolumn{1}{|c|}{$\mbox{ }$}      & 200  & 1030516/107, \ 0.00  & 8697/100, \ 0.00 & 299.3/106, \ 0.00 \\
    \multicolumn{1}{|c|}{$\Rsc$}     & 1000 & 204171/107, \ 0.00  & 314.2/106, \ 0.00 & 116.5/106, \ 0.23 \\
    \multicolumn{1}{|c|}{$\mbox{ }$}      & 5000 & 40879/107, \ 0.00  & 95.88/107, \ 0.77 & 92.46/107, \ 0.84 \\
    \multicolumn{1}{|c|}{$\mbox{ }$}      & 25000 & 8100/107, \ 0.00 & 107.2/107, \ 0.48 & 108.5/107, \ 0.44 \\
  \bottomrule
  \end{tabular}
\end{table}

%%%%%%%%%%%%%%%%%%%%%%%%%%%%%%%%%%%%%%%%%%%%%%%%%%%%%%%%%%%%%%%%%%%%%%%%%%%%%%%%%%%%%%
\end{document}